\documentclass[lettersize,journal]{IEEEtran}
\usepackage{amsmath,amsfonts}
\usepackage{algorithmic}
\usepackage{algorithm}
\usepackage{array}
\usepackage[caption=false,font=normalsize,labelfont=sf,textfont=sf]{subfig}
\usepackage{textcomp}
\usepackage{stfloats}
\usepackage{url}
\usepackage{verbatim}
\usepackage{graphicx}
\usepackage{cite}
\usepackage{bm}
\usepackage{makecell}
\usepackage{amssymb}
\usepackage{color}
\makeatletter
\renewcommand{\citepunct}{,\penalty\@m\hskip.13emplus.1emminus.1em}
\renewcommand{\citedash}{\hbox{--}\penalty\@m}

\begin{document}

\title{Gradient-Driven Graph Neural Networks for Learning Digital and Hybrid Precoder}

\author{Lin Zhang, Shengqian Han,
and Chenyang Yang
\thanks{The authors are with the School of Electronics and Information Engineering, Beihang University, Beijing 100191, China (e-mail:\{lin.zhang,sqhan,cyyang\}@buaa.edu.cn).}
\vspace{-0.4cm}
}



\maketitle

\begin{abstract}
The optimization of multi-user multi-input multi-output (MU-MIMO) precoders is a widely recognized challenging problem. Existing work has demonstrated the potential of graph neural networks (GNNs) in learning precoding policies. However, existing GNNs often exhibit poor generalizability for the numbers of users or antennas.
In this paper, we develop a gradient-driven GNN design method for the learning of fully digital and hybrid precoding policies. The proposed GNNs leverage two kinds of knowledge, namely the gradient of signal-to-interference-plus-noise ratio (SINR) to the precoders and the permutation equivariant property of the precoding policy. To demonstrate the flexibility of the proposed method for accommodating different optimization objectives and different precoding policies, we first apply the proposed method to learn the fully digital precoding policies. We study two precoder optimization problems for spectral efficiency (SE) maximization and log-SE maximization to achieve proportional fairness.
We then apply the proposed method to learn the hybrid precoding policy, where the gradients to analog and digital precoders are exploited for the design of the GNN.
Simulation results show the effectiveness of the proposed methods for learning different precoding policies and better generalization performance to the numbers of both users and antennas compared to baseline GNNs.
\end{abstract}


\section{Introduction}
\IEEEPARstart The optimization of multi-user multi-input multi-output (MU-MIMO) precoding is challenging and often requires iterative algorithms, such as the weighted minimum mean square error (WMMSE) algorithm~\cite{WMMSE}, to achieve superior performance. The iterative computations result in high computational demands, particularly as the number of users or antennas increases, making it difficult to meet the real-time requirements of mobile applications.

In recent years, deep neural networks (DNNs) have been introduced to learn complex wireless policies. Early work primarily employed the fully-connected neural networks (FNNs) or convolutional neural networks (CNNs) to learn the policies~\cite{DNN1,DNN2,DNN3,DNN6,DNN7}. While these methods can effectively reduce online inference time, they often require a large number of trainable parameters, massive data sets, and long training time \cite{DNN1,DNN2,DNN3,DNN6,DNN7}. Moreover, these DNNs lack \emph{size generalizability} for the learning of wireless policies, i.e., they have to be re-trained when the system scale changes, highlighting the need to reduce training complexity and improve generalizability~\cite{SCJ}.

Incorporating the mathematical properties of the policies to be learned into the design of DNNs has been recognized as a promising approach to improve learning efficiency by reducing the hypothesis space of the DNNs~\cite{ZBC,SCJ}. Graph neural networks (GNNs) can leverage the permutation equivariant (PE) property existed in many wireless policies \cite{SCJ,PE}, which has been exploited for learning tasks such as power allocation \cite{PE,GJ} and precoding policies~\cite{ZBC, LSJ, MDGNN}. Compared to FNNs and CNNs, GNNs incorporate parameter-sharing structures that satisfy the PE properties, leading to the advantages of improved learning performance, reduced training sample requirement, fewer learnable parameters, and decreased runtime. 

Another approach to enhance learning performance is integrating mathematical models with DNNs. This can be achieved by emulating existing iterative algorithms \cite{WMMSE1, WMMSE2, WMMSE3, WMMSE4, WMMSE5}, leveraging optimal solution structures~\cite{optimal1, optimal2, optimal3, optimal4, optimal5}, exploiting the formula of Shannon capacity \cite{DR}, or incorporating other domain-specific knowledge \cite{model}. Deep unfolding, a popular model-driven learning approach, uses DNNs to learn specific operations or parameters within iterative algorithms \cite{WMMSE1,WMMSE2, WMMSE3, WMMSE4, WMMSE5}. 


Deep learning has been applied to optimize precoding in different multi-antenna systems, which can significantly reduce inference time \cite{DNN2}, jointly optimize precoders with channel acquisition \cite{DNN3, DNN5}, or enhance robustness against imperfect channels \cite{DNN6}. To improve learning and generalization performance while reducing training complexity, existing studies have explored the integration of mathematical properties or mathematical models into DNN designs, which are reviewed~below.

\subsection{Related Works}

\subsubsection{Incorporating Mathematical Properties}
Leveraging the PE property of precoding policies can reduce the hypothesis space of DNNs and hence improve learning efficiency.
In \cite{ZBC}, an edge-GNN was designed to exploit the PE property,  and its advantages over CNNs were analyzed. In \cite{DNN4}, a vertex-GNN was designed to learn beamforming and reflective coefficients in intelligent reflecting surface-aided multi-antenna systems, where the GNN was trained to maximize the spectrum efficiency (SE) and the minimal data rate of users, respectively. In \cite{MDGNN}, a unified framework for designing multidimensional GNNs to learn permutable wireless policies was proposed, with the analog-digital hybrid precoder as an illustrative example.

Although the GNN in \cite{ZBC} leveraged the PE properties, their learning efficiency and performance degrade when the signal-to-noise ratio (SNR) is high or when the system scale (i.e., the number of users or antennas) increases. Furthermore, while GNNs such as the vertex-GNN in \cite{DNN4} and the edge-GNN in \cite{ZBC, MDGNN} have potential for generalization to larger system scales, their actual generalization performance remains limited.

\subsubsection{Incorporating Mathematical Models}
Deep unfolding is a widely-used model-driven learning method that uses the layers of a DNN to replicate the iteration process of an algorithm. It learns specific operations or parameters within iterative algorithms, or introduces additional learnable parameters into each iteration. In \cite{WMMSE1}, the WMMSE algorithm was unfolded to optimize precoding, aiming to reduce the complexity of the involved matrix inverse operations. In \cite{WMMSE2}, deep unfolding was applied to optimize power allocation, accelerating the algorithm's convergence. In \cite{WMMSE3}, deep graph unfolding was applied for the optimization of beamforming in MU-MIMO single-hop wireless ad-hoc interference networks. In \cite{WMMSE4}, the WMMSE algorithm was unfolded by graph convolutional network (GCN) to optimize coordinated beamforming, aiming to reduce the number of iterations. In \cite{WMMSE5}, deep unfolding was applied by treating important hyper-parameters of the iterative algorithm as learnable parameters for joint precoding and pilot design to reduce the complexity. Except for precoding, deep unfolding is also widely used for MIMO detection.
In \cite{DetNet}, signal detector was optimized by unfolding a projected gradient descent algorithm, where only a few parameters such as step size were learned from the data. In \cite{OAMPNet}, an orthogonal approximate message passing algorithm was unfolded by incorporating additional training parameters. In \cite{EPGNN} and \cite{AMPGNN}, signal detection algorithms, such as expectation propagation and approximate message passing, were unfolded using GNNs, respectively, to enhance the detection performance. Compared to iterative algorithms, deep unfolding networks can accelerate convergence and reduce inference complexity. However, since deep unfolding networks still involve numerical computations inherited from the original algorithms, their inference complexity remains higher than that of DNNs, making real-time implementation in large-scale wireless systems challenging. Moreover, the algorithm- and problem-specific design limits their flexibility and adaptability to diverse scenarios. 

Exploiting the structure of the optimal policy is another model-driven approach, which can simplify the policy to be learned. In \cite{optimal1, optimal2, optimal3, optimal4, optimal5}, with the structure of the optimal precoder, the learning of precoders is simplified to the learning of power allocations. While this approach improves learning performance and reduces training complexity, the works in \cite{optimal1, optimal2, optimal3, optimal4} do not leverage the PE properties, preventing them from being generalized to different system scales. Moreover, this approach is highly dependent on the specific structure of the optimal solution, making it problem-specific. For example, the optimal precoding solution structure used in \cite{optimal1, optimal2, optimal3, optimal4, optimal5} is derived in multi-user multiple-input single-output (MU-MISO) systems and cannot be extended to MU-MIMO systems. Therefore, the applicability of DNNs based on the optimal solution structure is significantly constrained.

Mathematical formulas can also aid in the design of DNNs. In \cite{catDHD}, a GNN was developed by using Taylor's expansion of the matrix pseudo-inverse, improving learning performance and efficiency. However, the GNN was specifically tailored for optimizing fully digital precoders and is not suitable for hybrid precoding policies. Moreover, to enhance the generalization performance to the number of users and antennas, the GNN needs to be trained on data sets consisting of different numbers of users and antennas, which are randomly sampled from an exponential distribution.

\subsection{Contributions}

In this paper, we propose a novel gradient-driven GNN design method for optimizing precoders, inspired by the correspondence between the gradient descent iteration equation and the GNN update equation. 
The major contributions are summarized as follows.\footnote{A part of this work, specifically the learning of the fully digital precoding policy for SE maximization, was presented in a conference paper~\cite{ZL-GCW23}. This journal version is substantially extended, including the learning of the fully digital precoding policy for achieving proportional fairness, the learning of the hybrid precoding policy, and new simulation results for both performance and complexity comparisons.}

\begin{itemize}
    \item[$\bullet$] We propose a gradient-driven GNN for learning the MU-MISO precoding policy. The design incorporates two kinds of knowledge: the gradient of the signal-to-interference-plus-noise ratio (SINR) with respect to the precoder, and the PE property of the precoding policies. The gradient information is exploited to design a novel information aggregation method, requiring only the aggregation of hidden representations from neighbor user vertices. Furthermore, it inspires the design of an attention mechanism to learn the aggregation weights, which significantly outperforms existing graph attention network (GAT) in  \cite{GAT} for precoder learning. We demonstrate the flexibility of the proposed method by applying it to learn the fully digital precoders for both SE and log-SE maximization. Simulation results show that the proposed GNN surpasses baseline GNNs in both learning and generalization performance.

    \item[$\bullet$] We show the versatility of the proposed gradient-driven GNN design method by applying it to learn the hybrid precoding policy. Drawing inspiration from the gradients for digital and analog precoders, we design a cascaded DNN structure for learning hybrid precoding policy with SE maximization as a case study. The results show that the proposed GNN achieves performance close to traditional numerical algorithms while significantly reducing inference complexity, and exhibits superior generalizability to the numbers of both users and antennas compared to other GNN-based~methods.
\end{itemize}

The rest of the paper is organized as follows. In Sec. \ref{system model}, we introduce the system model and existing GNN architectures for precoder learning. In Sec. \ref{Grad-GNN for digital}, we propose the gradient-driven GNN design method and apply it to learn the fully digital precoder, aimed at maximizing SE and log-SE, respectively. In Sec. \ref{Grad-GNN for hybrid}, we further extend the proposed GNN to learn hybrid precoders. In Sec. \ref{simulation}, the learning performance, generalizability, and complexity are compared by simulations. Finally, conclusions are drawn in Sec. \ref{conclusion}.

\emph{Notations:}
$(\cdot)^T$, $(\cdot)^H\!$, and $\left\|\cdot\right\|_F$ denote transpose, Hermitian transpose, and Frobenius norm of a matrix, respectively. $|\cdot|$ denotes the magnitude of a complex value. ${(\bf X)}_{i,j}$ denotes the element in the $i$-th row and $j$-th column of matrix $\bf X$. $\otimes$ denotes Kronecker product.

\section{System Model}  \label{system model}
Consider the downlink transmission where a base station (BS) equipped with \(N\) antennas transmits to \(K\) users, each with a single antenna. We first focus on the optimization of the fully digital precoder, which will later be extended to hybrid precoding.

A general formulation of the fully digital precoder optimization problem can be expressed as
\begin{subequations}\label{P0: max sinr}
	\begin{align}
		\max_{{\bf V}} ~~& u(\gamma_1, \cdots, \gamma_K) \label{P0: msr-1} \\
		{\rm s.t.}\  ~~& s({\bf V}) \leq 0, \label{P0: msr-2}
	\end{align}
\end{subequations}
where $u(\cdot)$ is the objective function, which is a function of the SINRs of $K$ users, $s(\cdot)$ is the constraint function of the precoding matrix ${\bf V}$, ${{\bf V} = [{\bf v}_{1},\cdots,{\bf v}_{K}]} \in {\mathbb C}^{N \times K}$, and \({\bf v}_k \in \mathbb{C}^{N \times 1}\) is the precoding vector for user \(k\).

The SINR $\gamma_k$ is expressed as
\begin{equation} \label{eq: SINR-digital}
	\gamma_k = \frac{|{\bf h}_k^{\rm H} {\bf v}_k|^2}{\sum_{j=1, j \neq k}^K |{\bf h}_k^{\rm H} {\bf v}_j|^2 + \sigma_n^2},
\end{equation}
where \({\bf h}_k \in \mathbb{C}^{N \times 1}\) is the channel vector from the BS to user \(k\), and \(\sigma_n^2\) denotes the noise variance.

This general optimization problem can be tailored to address a series of specific precoder optimization problems. For illustrative purposes, we consider two widely studied and challenging problems: SE maximization and log-SE maximization.
\subsubsection{\underline{SE Maximization}}
This problem aims to maximize the sum SE of users subject to the total power constraint of the BS, which is a typical problem for enhancing the overall system throughput. The optimization problem is
\begin{subequations}\label{P1: max sum-rate}
	\begin{align}
		\max_{{\bf V}} ~~& \sum_{k=1}^{K} \log_2 \left(1 + \gamma_{k}\right) \label{P1: msr-1} \\
		{\rm s.t.}\  ~~& \|{\bf V}\|_F^2 \leq P^{\max}, \label{P1: msr-2}
	\end{align}
\end{subequations}
where $P^{\max}$ is the total transmit power of the BS.

\subsubsection{\underline{Log-SE Maximization}}
This problem aims to achieve proportional fairness, i.e., balance the sum rate and the fairness among users, by maximizing the logarithms of the SE of users. The optimization problem is
\begin{subequations}\label{P2: max log-sum-rate}
	\begin{align}
		\max_{{\bf V}} ~~& \sum_{k=1}^{K} \ln \log_2 \left(1 + \gamma_{k}\right) \label{P2: msr-1} \\
		{\rm s.t.}\  ~~& \|{\bf V}\|_F^2 \leq P^{\max}. \label{P2: msr-2}
	\end{align}
\end{subequations}

\subsection{PE Properties of Fully Digital Precoding Policies} \label{digital-PE}
The precoding policies for either of the two exemplary problems are mappings from the channel matrix ${{\bf H} = [{\bf h}_1,\cdots,{\bf h}_K]} \in {\mathbb C}^{N \times K}$ to the optimized precoders ${\bf V}^\star$. Let ${\bf V}^{\star}=f_b({\bf H})$ denote the policy.
As analyzed in \cite{ZBC}, such a policy satisfies two-dimensional PE properties. In particular, let ${\bf\Pi}_{\text{U}}$ and ${\bf\Pi}_{\text{A}}$ denote two permutation matrices with respect to user indices and BS antenna indices, respectively. Then, the following property holds
\begin{equation} \label{eq: PE}
	{\bf\Pi}_{\text{A}}^T {\bf V}^\star {\bf\Pi}_{\text{U}} = f_b({\bf\Pi}_{\text{A}}^T {\bf H} {\bf\Pi}_{\text{U}}).
\end{equation}
It indicates that when the order of users or antennas changes, the optimal precoder changes accordingly while the SINR of each user does not change, such that the values of the objective functions of problem \eqref{P1: max sum-rate} and \eqref{P2: max log-sum-rate} do not change.

This property can be leveraged to design the parameter sharing scheme for GNNs, whose effectiveness in improving learning performance and reducing training complexity has been validated in previous studies~\cite{ZBC,catDHD}.

\subsection{Recap of GNN}  \label{sub: GNN}
To demonstrate the novelty of the proposed GNN, we briefly review the main components of the GNN designed in \cite{ZBC} (called vanilla-GNN in the sequel).

\begin{figure}
	\centering
	\includegraphics[width=0.50\textwidth]{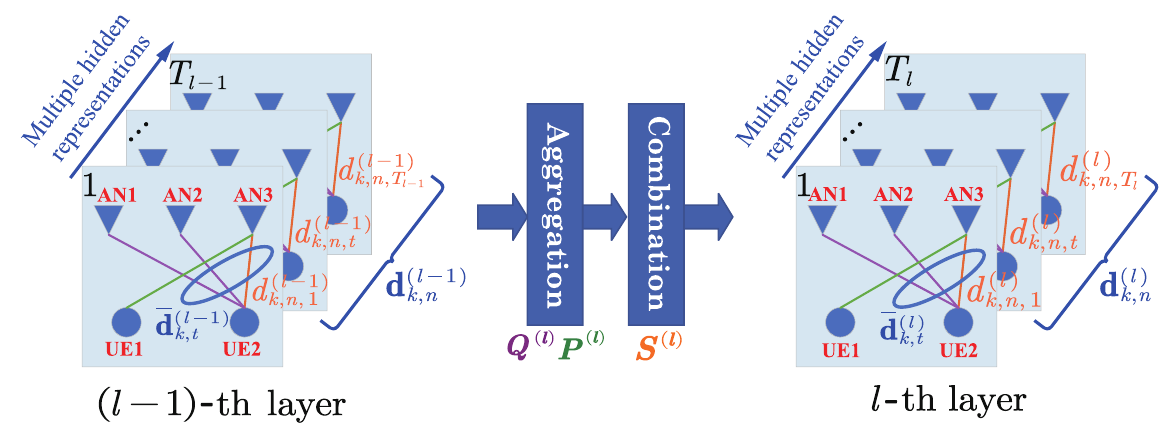}
	\caption{Graph and the iteration procedure of GNN for learning digital precoding policy.}
	\label{fig:GNN}
\end{figure}

GNNs learn over graphs. To solve problem \eqref{P1: max sum-rate}, a heterogeneous graph was built in \cite{ZBC}, as shown in Fig. \ref{fig:GNN}, where the BS antennas and users were defined as two types of vertices and the edges exist between the two types of vertices. Each vertex does not have features, and the feature of edge $(k,n)$ is the channel coefficient $h_{k,n}$ from antenna $n$ to user $k$. The actions are defined on edges, which constitute the learned precoding~matrix.

GNN iteratively performs aggregation and combination steps, with which the information is propagated throughout the entire graph and the complex relationship between the input and output variables can be captured. Since both the features and actions are defined on edges, the GNN in \cite{ZBC} updates the hidden representations of edges, where the aggregation and combination steps can be described as follows.
\begin{itemize}
	\item[(i)] \textbf{Aggregation}: In the $l$-th layer, edge $(k,n)$ aggregates information from its neighbor edges, where neighbor edges are the ones connected to user vertex $k$ or antenna vertex $n$. The hidden representations of the edges in the $l$-th layer form a tensor with the dimension $K\times N\times T_l$, where $T_l$ is the number of hidden representations. For edge $(k,n)$, the $T_l$ hidden representations are denoted by vector ${\mathbf{d}}_{k,n}^{(l)} = [d_{k,n,1}^{(l)}, \dots, d_{k,n,T_{l}}^{(l)}]^T$.
Then, from the neighbor edges connected to user vertex $k$, denoted by set $\mathcal{N}_{{\rm{u}},k}$, the aggregated information can be expressed as
\begin{align}\label{eq: gnn aggregator10}
			{\bf u}_{k,n}^{(l)}
			\!=\! {\sf PL}_{u}\Big(q_u ({\bf d}_{k,i}^{(l-1)},{\bf Q}^{(l)})|i\in\mathcal{N}_{{\rm{u}},k}\Big),
\end{align}
where ${\sf PL}_{u}(\cdot)$ denotes the pooling function, and $q_u(\cdot,{\bf Q}^{(l)})$ is the processing function with the learnable parameter ${\bf Q}^{(l)}\in\mathbb{R}^{T_l\times T_{l-1}}$. The aggregated information from the neighbor edges connected to antenna vertex $n$ can be expressed as
\begin{align}\label{eq: gnn aggregator20}
			{\bf a}_{k,n}^{(l)}
			\!=\! {\sf PL}_{a}\Big(q_a ({\bf d}_{j,n}^{(l-1)},{\bf P}^{(l)})|j\in\mathcal{N}_{{\rm{a}},n}\Big),
\end{align}
where $\mathcal{N}_{{\rm{a}},n}$, ${\sf PL}_{a}(\cdot)$, $q_a(\cdot,{\bf P}^{(l)})$, and ${\bf P}^{(l)}$ are defined similar to those in \eqref{eq: gnn aggregator10}.
	\item[(ii)] \textbf{Combination}: The aggregated information from neighbor edges is combined with the hidden representations of edge $(k,n)$ in the $(l-1)$-th layer as
	\begin{equation} \label{eq: gnn combiner}
		{\bf d}_{k,n}^{(l)} = {\sf CB}\Big({\bf d}_{k,n}^{(l-1)}, {\bf u}_{k,n}^{(l)}, {\bf a}_{k,n}^{(l)}, {\bf S}^{(l)}\Big),
	\end{equation}
	where ${\sf CB}(\cdot,{\bf S}^{(l)})$ denotes the combination function with the learnable parameter ${\bf S}^{(l)}\in\mathbb{R}^{T_l\times T_{l-1}}$.
\end{itemize}

The pooling, processing, and combination functions should be selected to make the GNN satisfy the PE properties in \eqref{eq: PE}. A common setup in the literature is to use linear functions for $q_u(\cdot)$ and $q_a(\cdot)$, summation/mean/maximum/minimum functions for ${\sf PL}_{u}(\cdot)$ and ${\sf PL}_{a}(\cdot)$, and linear functions cascaded with an activation function $\sigma (\cdot)$ for ${\rm CB}(\cdot)$. In particular, the vanilla-GNN in \cite{ZBC} was designed with the following update equation.
	
    \begin{equation}  \label{eq: gnn combiner1}
        \begin{aligned}
		&{\bf d}_{k,n}^{(l)} = \sigma\left({\bf S}^{(l)}{\bf d}_{k,n}^{(l-1)}+{\bf u}_{k,n}^{(l)}+{\bf a}_{k,n}^{(l)}\right) \\
        &= \!\sigma\!\left(\!{\bf S}^{(l)}{\bf d}_{k,n}^{(l-1)}\!+\!\sum_{i\in\mathcal{N}_{{\rm{u}},k}} \!{\bf Q}^{(l)} {\bf d}_{k,i}^{(l-1)}\!+\!\sum_{j\in\mathcal{N}_{{\rm{a}},n}} \!{\bf P}^{(l)} {\bf d}_{j,n}^{(l-1)} \!\right).
        \end{aligned}
    \end{equation}

\section{Design of Gradient-Driven GNNs} \label{Grad-GNN for digital}
In this section, we examine the connection between the classic gradient descent algorithm and the update equation of GNN, based on which a novel GNN is developed. Then, we apply the designed GNN to solve the two exemplary problems \eqref{P1: max sum-rate} and \eqref{P2: max log-sum-rate}.

\subsection{Gradient Descent Algorithm}
Gradient descent is a generic method for continuous optimization. When using the gradient descent algorithm to optimize the precoder, the precoding vector of user $k$ in the $l$-th iteration can be computed as
\begin{equation} \label{eq: gradient descent iteration}
	{\bf v}_k^{(l)}={\bf v}_k^{(l-1)}-\delta \bigtriangledown_{{\bf v}_k}u(\gamma_1,\cdots,\gamma_K),
\end{equation}
where $\delta$ is the step size of each iteration, $u(\gamma_1,\cdots,\gamma_K)$ is the objective function of problem \eqref{P0: max sinr}, and $\bigtriangledown_{{\bf v}_k}u$ is the gradient of the objective function.

The gradient $\bigtriangledown_{{\bf v}_k}u$ can be obtained as
\begin{equation} \label{eq: gradient-V}
	\begin{aligned}
		& \bigtriangledown_{{\bf v}_k}u = \sum_{i=1}^{K} \frac {\partial u(\gamma_1,\cdots,\gamma_K)}{\partial \gamma_i} \cdot \frac{\partial \gamma_i}{\partial \bf v_k}  \\
		& =\underbrace{\frac {\partial u(\gamma_1,\cdots,\gamma_K)}{\partial \gamma_k} \cdot \frac{\partial \gamma_k}{\partial {\bf v}_k}}_{\triangleq \Delta_i({\bf h}_i,{\bf V}^{(l-1)}), i=k} + \sum_{i=1,i\neq k}^K \underbrace{\frac {\partial u(\gamma_1,\cdots,\gamma_K)}{\partial \gamma_i} \cdot \frac{\partial \gamma_i}{\partial {\bf v}_k}}_{\triangleq \overline{\Delta}_i({\bf h}_i,{\bf V}^{(l-1)}), i\neq k}.
	\end{aligned}
\end{equation}
Denoting the term $\frac {\partial u(\gamma_1,\cdots,\gamma_K)}{\partial \gamma_i}$ as $\lambda_i$, the terms $\Delta_i({\bf h}_i,{\bf V}^{(l-1)})$ and $\overline{\Delta}_i({\bf h}_i,{\bf V}^{(l-1)})$ can be obtained after some manipulations as
\begin{align}
	\Delta_i({\bf h}_i,{\bf V}^{(l-1)})&=\lambda_i \cdot \frac{{\bf h}_{i}^{\rm H}{\bf v}_{k}^{(l-1)}}{\sum_{j=1,j\neq i}^K |{\bf h}_{i}^{\rm H} {\bf v}_{j}^{(l-1)}|^2 + \sigma^2}{\bf h}_{i}\nonumber\\
    &\triangleq \rho_{k,i}({\bf h}_i,{\bf V}^{(l-1)}) {\bf h}_{i}, \ i= k.\label{eq: fk} \\
    \overline{\Delta}_i({\bf h}_i,{\bf V}^{(l-1)})&=\lambda_i \cdot \frac{-|{\bf h}_{i}^{\rm H}{\bf v}_{i}^{(l-1)}|^2 \cdot {\bf h}_{i}^{\rm H}{\bf v}_{k}^{(l-1)}}{\left(\sum_{j=1,j\neq i}^K |{\bf h}_{i}^{\rm H} {\bf v}_{j}^{(l-1)}|^2 + \sigma^2 \right)^2} {\bf h}_{i} \nonumber\\
    &\triangleq \rho_{k,i}({\bf h}_i,{\bf V}^{(l-1)}) {\bf h}_{i}, \ i\neq k.\label{eq: fi}
\end{align}

Upon substituting \eqref{eq: fk} and \eqref{eq: fi} into \eqref{eq: gradient-V} and then \eqref{eq: gradient descent iteration}, we can rewrite the gradient descent iteration equation as

\begin{equation} \label{eq: gradient descent iteration-V}
	{\bf v}_k^{(l)}={\bf v}_k^{(l-1)}-\delta\sum_{i=1}^K \rho_{k,i}({\bf h}_i,{\bf V}^{(l-1)}){\bf h}_{i},
\end{equation}
where the function $\rho_{k,i}({\bf h}_i,{\bf V}^{(l-1)})$ is defined in \eqref{eq: fk} and \eqref{eq: fi}.

While applying the gradient descent algorithm directly to optimize the precoder might not yield satisfactory performance, the derived iteration equation \eqref{eq: gradient descent iteration-V} can inspire the design of the update equation of GNN in the next~subsection.

\begin{figure}
	\centering
	\includegraphics[width=0.48\textwidth]{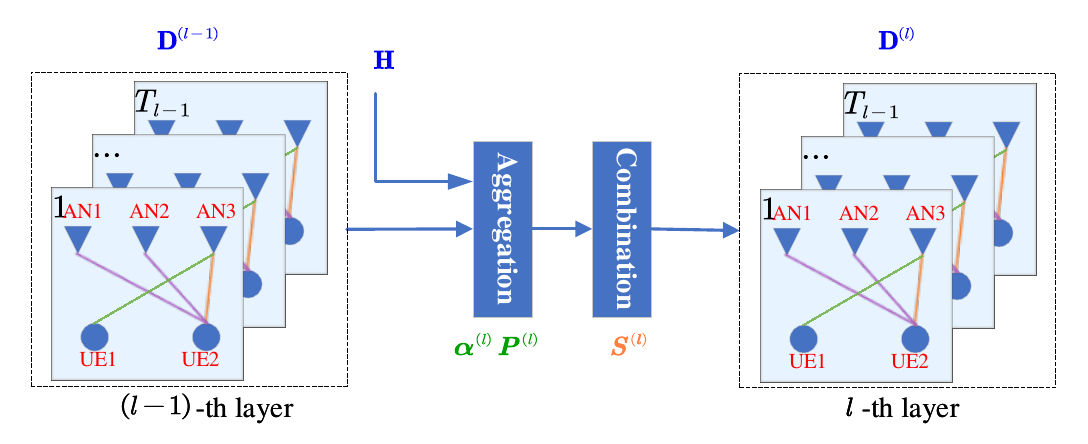}
	\vspace{-1.5mm}
	\caption{The iteration procedure of gradient-driven GNN for learning digital precoding policy.}
	\label{fig:full-digital}
\end{figure}

\subsection{Gradient-Driven GNN}
\subsubsection{\underline{GNN Design}}
By comparing \eqref{eq: gradient descent iteration-V} with \eqref{eq: gnn combiner1}, we can observe the similarity between the gradient descent iteration equation and the update equation of GNN. First, both involve iterations to compute the precoder ${\bf v}_k^{(l)}$ or hidden representation ${\bf d}_{k,n}^{(l)}$. Second, in each iteration, the precoder, say ${\bf d}_{k,n}^{(l-1)}$ in \eqref{eq: gnn combiner1} or ${\bf v}_k^{(l-1)}$ in \eqref{eq: gradient descent iteration-V}, is updated with the aggregated information, i.e., ${\bf u}_{k,n}^{(l)}$ and ${\bf a}_{k,n}^{(l)}$ in \eqref{eq: gnn combiner1} and $\rho_{k,i}({\bf h}_i,{\bf V}^{(l-1)}){\bf h}_{i}$ in \eqref{eq: gradient descent iteration-V}.

This inspires us to leverage the gradient descent iteration equation \eqref{eq: gradient descent iteration-V} to design the update equation of GNN. Compared to the vanilla-GNN given by \eqref{eq: gnn combiner1}, \eqref{eq: gradient descent iteration-V} has three major differences in information aggregation and combination.
\begin{itemize}
  \item[i)] \eqref{eq: gradient descent iteration-V} contains a single summation term with respect to users, implying that we only need to aggregate the information from the neighbor user vertices. By contrast, the vanilla-GNN requires aggregation from both neighbor user and antenna vertices.
  \item[ii)] In \eqref{eq: gradient descent iteration-V}, the summation aggregates the channel vector ${\bf h}$ from neighbor users, whereas \eqref{eq: gnn combiner1} aggregates the hidden representation ${\bf d}^{(l-1)}$ from neighbor users and antennas.
  \item[iii)] In \eqref{eq: gradient descent iteration-V}, the summation over the channel vectors assigns a unique weight $\rho_{k,i}({\bf h}_i,{\bf V}^{(l-1)})$ to each ${\bf h}_{i}$, while in \eqref{eq: gnn combiner1} all hidden representations ${\bf d}_{k,i}^{(l-1)}$ share the same weight ${\bf Q}^{(l)}$~or~${\bf P}^{(l)}$.
\end{itemize}

The three observations enable a new design of the aggregation and combination steps of GNN. For ease of elaboration, let us first consider that the edges in each layer have only a single hidden representation, say the $t$-th one.
The designed update equation of the GNN can be expressed as
\begin{equation} \label{eq: combination equation-V}
	{d}_{k,n,t}^{(l)}=\sigma\left(s^{(l)}{d}_{k,n,t}^{(l-1)}+p^{(l)}\sum_{i=1}^K \alpha_{k,i,t}^{(l-1)}h_{i,n}\right),
\end{equation}
where $s^{(l)}$ and $p^{(l)}$ are two learnable scalar parameters, and $\alpha_{k,i,t}^{(l-1)}$ is a learnable weight for information aggregation. We can find that \eqref{eq: combination equation-V} well matches the three observations. The proposed gradient-driven GNN for learning digital precoding is shown in Fig. \ref{fig:full-digital}.

The weight $\alpha_{k,i,t}^{(l-1)}$ reflects the importance of the information provided by user $i$ to user $k$, which falls into the framework of attention mechanisms. As in \cite{GAT}, we use a single-layer fully-connected neural network $\phi(\cdot)$ to learn the attention coefficient.  
Nevertheless, inspired by the expression of $\rho_{k,i}({\bf h}_i,{\bf V}^{(l-1)})$ in \eqref{eq: fk} and \eqref{eq: fi}, we consider a different input for $\phi(\cdot)$ from that in \cite{GAT}. Specifically, instead of using the concatenation of $\bar{\bf d}_{k,t}^{(l-1)}$ and $\bar{\bf d}_{j,t}^{(l-1)}$ as input, as done in \cite{GAT}, we define the attention mechanism and its input as
\begin{align} \label{eq: attention-NN-V}
	\alpha_{k,i,t}^{(l-1)}=&\phi\Big({\bf h}_{i}^{\rm H}\bar{\bf d}_{k,t}^{(l-1)},\ \textstyle\sum_{j=1,j\neq i}^K |{\bf h}_{i}^{\rm H}\bar{\bf d}_{j,t}^{(l-1)}|^2, \nonumber\\
&\qquad |{\bf h}_{i}^{\rm H}\bar{\bf d}_{i,t}^{(l-1)}|^2\cdot\xi_{i,k}\Big),
\end{align}
where $\bar{\bf d}_{k,t}^{(l-1)} = [d_{k,1,t}^{(l-1)}, \dots, d_{k,N_t,t}^{(l-1)}]^T$ denotes the $t$-th hidden representations from all antennas to user $k$, i.e., the precoding vector of user $k$. The network $\phi(\cdot)$ has three inputs, where the third input depends on a variable $\xi_{i,k}$, which is defined as
\begin{equation} \label{eq: theta}
	\xi_{i,k}=
	\begin{cases}
		0,&   i=k,\\
		1,&   i \neq k.
	\end{cases}
\end{equation}
The use of $\xi_{i,k}$  comes from the expression of $\rho_{k,i}({\bf h}_i,{\bf V}^{(l-1)})$. When $i=k$, we can find from \eqref{eq: fk} that $\rho_{k,k}({\bf h}_k,{\bf V}^{(l-1)})$ depends on the terms ${\bf h}_{k}^{\rm H}{\bf v}_{k}^{(l-1)}$ and $\sum_{j=1, j\neq i}^K |{\bf h}_{k}^{\rm H} {\bf v}_{j}^{(l-1)}|^2$, which make up the first two inputs of $\phi(\cdot)$. When $i\neq k$, as shown by \eqref{eq: fi}, $\rho_{k,i}({\bf h}_i,{\bf V}^{(l-1)})$ also depends on the term $|{\bf h}_{i}^{\rm H} {\bf v}_{i}^{(l-1)}|^2$, which is the third input of $\phi(\cdot)$.

\subsubsection{\underline{PE Properties}} \label{PE-digital}
We next show that the proposed GNN satisfies the PE property in \eqref{eq: PE}.
\begin{itemize}
  \item[i)] PE Property for Antenna Indices: Assume that in the $l$-th layer the antenna index $n$ of the input ${d}_{k,n,t}^{(l-1)}$ and ${h}_{i,n,t}^{(l-1)}$ is permuted to $m$, then the right-hand side of \eqref{eq: combination equation-V} becomes
  \begin{align} \label{E:antennaPE}
    \sigma\left(s^{(l)}{d}_{k,m,t}^{(l-1)}+p^{(l)}\sum_{i=1}^K \alpha_{k,i,t}^{(l-1)}h_{i,m}\right).
  \end{align}
  Since the permutation of antennas does not affect the value of $\alpha_{k,i,t}^{(l-1)}$ as shown by \eqref{eq: attention-NN-V}, we can readily find from \eqref{eq: combination equation-V} that \eqref{E:antennaPE} equals ${d}_{k,m,t}^{(l)}$, i.e., the output of the $l$-th layer permutes the index accordingly. Thus, the PE property holds for the antenna indices.
  \item[ii)] PE Property for User Indices: Assume that in the $l$-th layer the user index $k$ of the input ${d}_{k,n,t}^{(l-1)}$ and ${h}_{k,n,t}^{(l-1)}$ is permuted to $c$.
  Since the attention mechanism $\phi(\cdot)$ given by \eqref{eq: attention-NN-V} is applied to all users, we know that permuting user index $k$ to $c$ will make $\alpha_{k,i,t}^{(l-1)}$ become $\alpha_{c,i,t}^{(l-1)}$.
  Thus, the right-hand side of \eqref{eq: combination equation-V} becomes
  \begin{align} \label{E:UEPE}
    \sigma\left(s^{(l)}{d}_{c,n,t}^{(l-1)}+p^{(l)}\sum_{i=1}^K \alpha_{c,i,t}^{(l-1)}h_{i,n}\right),
  \end{align}
  which equals ${d}_{c,n,t}^{(l)}$, indicating that the PE property holds for the user indices.
\end{itemize}

\subsubsection{\underline{Extension to Multiple Hidden Representations}}$\!\!\!$Consider that in the $(l-1)$-th and $l$-th layers of GNN, the edges have $T_{l-1}$ and $T_{l}$ hidden representations, respectively. Then, we can extend the update equation given by \eqref{eq: combination equation-V} as
\begin{equation} \label{eq: combination equation}
	{\bf d}_{k,n}^{(l)}=\sigma\left({\bf S}^{(l)}{\bf d}_{k,n}^{(l-1)}+{\bf P}^{(l)}\sum_{i=1}^K {\boldsymbol \alpha}_{k,i}^{(l-1)} h_{i,n} \right),
\end{equation}
where ${\mathbf{d}}_{k,n}^{(l)} = [d_{k,n,1}^{(l)}, \dots, d_{k,n,T_{l}}^{(l)}]^T$, ${\mathbf{d}}_{k,n}^{(l-1)} = [d_{k,n,1}^{(l-1)}, \dots, d_{k,n,T_{l-1}}^{(l-1)}]^T$, ${\bf S}^{(l)}, {\bf P}^{(l)}\in\mathbb{R}^{T_l\times T_{l-1}}$ are the learnable matrices, and ${\boldsymbol \alpha}_{k,i}^{(l-1)} = [\alpha_{k,i,1}^{(l-1)}, \dots, \alpha_{k,i,T_{l-1}}^{(l-1)}]^T$ is the attention weight vector.

To obtain ${\boldsymbol \alpha}_{k,i}^{(l-1)}$, we extend the input and output dimensions of the attention mechanism given by \eqref{eq: attention-NN-V} as
\begin{align} \label{eq: attention-NN1}
	& {\boldsymbol \alpha}_{k,i}^{(l-1)}=\phi\bigg(\left[{\bf h}_{i}^{\rm H}\bar{\bf d}_{k,t}^{(l-1)},\dots,{\bf h}_{i}^{\rm H}\bar{\bf d}_{k,T_{l-1}}^{(l-1)}\right], \nonumber\\
&\quad \  \left [\textstyle\sum_{j=1,j\neq i}^K |{\bf h}_{i}^{\rm H}\bar{\bf d}_{j,1}^{(l-1)}|^2, \dots, \textstyle\sum_{j=1,j\neq i}^K |{\bf h}_{i}^{\rm H}\bar{\bf d}_{j,T_{l-1}}^{(l-1)}|^2\right], \nonumber\\
&\quad\ \left[|{\bf h}_{i}^{\rm H}\bar{\bf d}_{i,1}^{(l-1)}|^2, \dots, |{\bf h}_{i}^{\rm H}\bar{\bf d}_{i,T_{l-1}}^{(l-1)}|^2\right]\cdot\xi_{i,k}\bigg).
\end{align}

We refer to the GNN with the update equation in \eqref{eq: combination equation} as gradient-driven GNN.

\label{extension-digital}

\vspace{-0.3cm}

\subsection{Learning Fully Digital Precoders with the Proposed GNN}
The proposed gradient-driven GNN is designed based on the gradient expression \(\bigtriangledown_{{\bf v}_k} u = \sum_{i=1}^{K} \frac{\partial u(\gamma_1, \cdots, \gamma_K)}{\partial \gamma_i} \cdot \frac{\partial \gamma_i}{\partial {\bf v}_k} = \sum_{i=1}^{K} \lambda_i \cdot \frac{\partial \gamma_i}{\partial {\bf v}_k}\), where the objective function only affects the first term, i.e., \(\lambda_i\). This scalar \(\lambda_i\) does not affect the update equation of the proposed GNN as shown in \eqref{eq: combination equation} and \eqref{eq: attention-NN1}. Therefore, the proposed gradient-driven GNN can be flexibly applied to optimize the precoder for different objective functions related to the SINR $\gamma_k$. In this subsection, we employ the proposed GNN to learn the precoders for the two problems, \eqref{P1: max sum-rate} and \eqref{P2: max log-sum-rate}, respectively.

\subsubsection{\underline{SE Maximization}}
Denote the objective function of problem \eqref{P1: max sum-rate} by \( g(\gamma_1,\cdots,\gamma_K) \). The gradient of the objective function with respect to \({\bf v}_k\) can be expressed as
\begin{equation} \label{gradient-SE}
    \bigtriangledown_{{\bf v}_k}g=\sum_{i=1}^{K} \frac {\partial g(\gamma_1,\cdots,\gamma_K)}{\partial \gamma_i} \cdot \frac{\partial \gamma_i}{\partial {\bf v}_k}.
\end{equation}
The term $\frac {\partial g(\gamma_1,\cdots,\gamma_K)}{\partial \gamma_i}=\frac {1}{1 + \gamma_i}$, which is a scalar that does not change the update equation of the proposed GNN. Hence, we can use the proposed GNN to learn the solution of the SE maximization problem. Specifically, the input and output of gradient-driven GNN are real and imaginary parts of channel and precoding, respectively, (i.e., ${\bf d}_{k,n}^{(0)}=[{\rm Re}(h_{k,n}), {\rm Im}(h_{k,n})]$ and ${\bf d}_{k,n}^{(L)}=[{\rm Re}(v_{k,n}), {\rm Im}(v_{k,n})]$). We train the gradient-driven GNN by employing the negative sum SE averaged over all training samples as the loss function, and the power constraint can be satisfied by passing the output of the GNN through a normalization layer, i.e., $\sqrt{P_{max}} \frac{\bf V}{||{\bf V}||_F}$.

\subsubsection{\underline{Log-SE Maximization}}
Problem \eqref{P2: max log-sum-rate} introduces a different objective function from problem \eqref{P1: max sum-rate} to ensure fairness among users. Despite this change, the proposed GNN is still suitable to solve the problem.

Denote the objective function of problem \eqref{P2: max log-sum-rate} by \( \eta(g(\gamma_1,\cdots,\gamma_K)) \). The gradient of the objective function with respect to \({\bf v}_k\) can be expressed as
\begin{equation} \label{gradient-PF}
    \bigtriangledown_{{\bf v}_k}\eta=\sum_{i=1}^{K} \frac {\partial \eta(g(\gamma_1,\cdots,\gamma_K))}{\partial g(\gamma_1,\cdots,\gamma_K)} \cdot \frac {\partial g(\gamma_1,\cdots,\gamma_K)}{\partial \gamma_i} \cdot \frac{\partial \gamma_i}{\partial {\bf v}_{k}}.
\end{equation}
Compared to the gradient in \eqref{gradient-SE}, the gradient in \eqref{gradient-PF} includes an additional scalar term, \(\frac {\partial \eta(g(\gamma_1,\cdots,\gamma_K))}{\partial g(\gamma_1,\cdots,\gamma_K)} = \frac {1}{g(\gamma_1,\cdots,\gamma_K)}\). Again, this term does not affect the update equation of the proposed GNN. Therefore, we can employ the designed gradient-driven GNN to learn the precoder for problem \eqref{P2: max log-sum-rate}. Compared to the learning for problem \eqref{P1: max sum-rate}, we only need to change the loss function during training to the negative logarithmic SE, i.e., the negative of the objective function in~\eqref{P2: max log-sum-rate}.

\section{Gradient-Driven GNN for Learning\\ Hybrid Precoding} \label{Grad-GNN for hybrid}
To further demonstrate the versatility of the proposed gradient-driven GNN design method, we extend its application to the learning of hybrid precoder in this section. Unlike the previous section that only learns the digital precoder, here we jointly learn both analog and digital precoders. To design the gradient-driven GNN for the problem, we first derive the gradients of analog and digital precoders, which will form the basis for the new GNN.

\subsection{Gradient Descent Algorithm for Hybrid Precoding}

Consider the downlink transmission with hybrid precoding, where the BS is equipped with $N_s$ RF chains, and $N_s$ is smaller than the number of antennas $N$. The optimization problem of hybrid precoders can be formulated as
\begin{subequations}\label{P3: max sum-rate-hybrid}
	\begin{align}
		\max_{{\bf W}_{\text{A}},{\bf W}_{\text{D}}} ~~& r(\gamma_1,\cdots,\gamma_K) \label{P3: msr-1} \\
		{\rm s.t.}\  ~~& |({\bf W}_{\text{A}})_{j,l}|=1,j=1,\cdots,N,l=1,\cdots,N_s, \label{P3: msr-2} \\
        ~~& s({\bf W}_{\text{A}},{\bf W}_{\text{D}}) \leq 0,
        \label{P3: msr-3}
	\end{align}
\end{subequations}
where ${\bf W}_{\text{A}}\in {\mathbb C}^{N \times N_s}$ is the analog precoding matrix, ${{\bf W}_{\text{D}} = [{\bf w}_{\text{D}1},\cdots,{\bf w}_{\text{D}K}]} \in {\mathbb C}^{N_s \times K}$ is the digital precoding matrix, ${\bf w}_{\text{D}k}\in \mathbb{C}^{N_s \times 1}$ is the digital precoder for user $k$, \eqref{P3: msr-2} is the constant modulus constraint for the analog precoder, and \eqref{P3: msr-3} denotes other constraints, say the power constraint of the BS. The objective function $r(\gamma_1,\cdots,\gamma_K)$ in \eqref{P3: msr-1} is related to the SINR of user $k$, $\gamma_{k}$, $i=1,\dots,K$, which is
\begin{equation} \label{eq: SINR-hybrid}
	\gamma_{k} = \frac{|{\bf h}_{k}^{\rm H} {\bf W}_{\text{A}} {\bf w}_{\text{D}k}|^2}{\sum_{j=1,j\neq k}^K |{\bf h}_{k}^{\rm H} {\bf W}_{\text{A}} {\bf w}_{\text{D}j}|^2 + \sigma_n^2}.
\end{equation}

Denote the hybrid precoding policy as $\{{\bf W}_{\text{A}}^\star,{\bf W}_{\text{D}}^\star\}=f_h({\bf H})$, where ${\bf W}_{\text{A}}^\star$ and ${\bf W}_{\text{D}}^\star$ are the optimal analog and digital precoding matrices, and $f_h(\cdot)$ is a multivariate function. The policy satisfies a three-set permutation property~\cite{MDGNN}
\begin{equation} \label{eq: hybrid-PE}
	\{{\bf\Pi}_{\text{A}}^T {\bf W}_{\text{A}}^\star {\bf\Pi}_{\text{R}}, {\bf\Pi}_{\text{R}}^T {\bf W}_{\text{D}}^\star {\bf\Pi}_{\text{U}} \} = f_h({\bf\Pi}_{\text{A}}^T {\bf H} {\bf\Pi}_{\text{U}}),
\end{equation}
where ${\bf\Pi}_{\text{A}}$, ${\bf\Pi}_{\text{R}}$ and ${\bf\Pi}_{\text{U}}$ are arbitrary permutation matrices that change the indices of antennas, RF chains, and users. Compared to the fully digital precoding policy, which satisfies the 2D-PE property given by \eqref{eq: PE}, we can find from \eqref{eq: hybrid-PE} that the hybrid precoding policy is equivariant to the permutation of user and antenna indices, and is invariant to the permutation of RF chain indices. Hence, the parameter sharing to satisfy the PE property of the policy differs from the GNN for learning the fully digital precoding policy.

To design the gradient-driven GNN for learning hybrid precoders, we first derive the gradient descent iteration equations for digital and analog precoders, respectively, which are expressed as
\begin{equation} \label{eq: gradient descent iteration-hybrid-Wbb}
	{\bf w}_{\text{D}k}^{(l)}={\bf w}_{\text{D}k}^{(l-1)}-\delta \bigtriangledown_{{\bf w}_{\text{D}k}}r(\gamma_1,\cdots,\gamma_K),
\end{equation}
\begin{equation} \label{eq: gradient descent iteration-hybrid-Wrf}
	\vec{\bf W}_{\text{A}}^{(l)}=\vec{\bf W}_{\text{A}}^{(l-1)}-\delta \bigtriangledown_{\vec{\bf W}_{\text{A}}}r(\gamma_1,\cdots,\gamma_K),
\end{equation}
where $r(\gamma_1,\cdots,\gamma_K)$ is the objective function in \eqref{P3: msr-1}, $\bigtriangledown_{{\bf w}_{\text{D}k}}r$ and $\bigtriangledown_{\vec{\bf W}_{\text{A}}}r$ are the gradients of the objective function for digital and analog precoders, respectively, and $\vec{\bf W}_{\text{A}}$ denotes the vectorization of ${\bf W}_{\text{A}}$.

For digital precoder, denoting the analog precoded channel ${\bf W}_{\text{A}}^H{\bf h}_i$ as $\hat{\bf h}_i\in {\mathbb C}^{N_s \times 1}$, the gradient of the objective function with respect to the digital precoder can be obtained as
\begin{equation} \label{eq: gradient-Wbb}
	\begin{aligned}
		\bigtriangledown_{{\bf w}_{\text{D}k}}r&
            =\sum_{i=1}^K \frac {\partial r(\gamma_1,\cdots,\gamma_K)}{\partial \gamma_i} \cdot \frac{\partial \gamma_i}{\partial {\bf w}_{\text{D}k}} \\
		&=\underbrace{\frac {\partial r(\gamma_1,\cdots,\gamma_K)}{\partial \gamma_k} \cdot \frac{\partial \gamma_k}{\partial {\bf w}_{\text{D}k}}}_{\triangleq \Delta_i(\hat{\bf h}_i,{\bf W}_{\text{D}}^{(l-1)}), i=k}  \\
      &+ \sum_{i=1,i\neq k}^K \underbrace{\frac {\partial r(\gamma_1,\cdots,\gamma_K)}{\partial \gamma_i} \cdot \frac{\partial \gamma_i}{\partial {\bf w}_{\text{D}k}}}_{\triangleq \overline{\Delta}_i(\hat{\bf h}_i,{\bf W}_{\text{D}}^{(l-1)}), i\neq k}.
	\end{aligned}
\end{equation}
Denoting $\frac {\partial r(\gamma_1,\cdots,\gamma_K)}{\partial \gamma_i}$ as $\lambda_i$, the terms $\Delta_i(\hat{\bf h}_i,{\bf W}_{\text{D}}^{(l-1)})$ and $\overline{\Delta}_i(\hat{\bf h}_i,{\bf W}_{\text{D}}^{(l-1)})$ can be obtained after some manipulations as
\begin{align}
	\Delta_i(\hat{\bf h}_i,{\bf W}_{\text{D}}^{(l-1)})&=\lambda_i\cdot\frac{\hat{\bf h}_{i}^{\rm H}{\bf w}_{\text{D}k}^{(l-1)}}{\sum_{j=1,j\neq i}^K |\hat{\bf h}_{i}^{\rm H} {\bf w}_{\text{D}j}^{(l-1)}|^2 + \sigma^2}\hat{\bf h}_{i}\nonumber\\
    &\triangleq \rho_{k,i}^{\text{D}}(\hat{\bf h}_i,{\bf W}_{\text{D}}^{(l-1)}) \hat{\bf h}_{i}, \ i= k.\label{eq: Wbb-fk}
\end{align}
\begin{align}
    \overline{\Delta}_i(\hat{\bf h}_i,{\bf W}_{\text{D}}^{(l-1)})&=\!\lambda_i\!\cdot\!\frac{-|\hat{\bf h}_{i}^{\rm H}{\bf w}_{\text{D}i}^{(l-1)}|^2 \cdot \hat{\bf h}_{i}^{\rm H}{\bf w}_{\text{D}k}^{(l-1)}}{\left (\sum_{j=1,j\neq i}^K |\hat{\bf h}_{i}^{\rm H} {\bf w}_{\text{D}j}^{(l-1)}|^2 \!+ \!\sigma^2 \right)^2} \hat{\bf h}_{i} \nonumber\\
    &\triangleq \rho_{k,i}^{\text{D}}(\hat{\bf h}_i,{\bf W}_{\text{D}}^{(l-1)}) \hat{\bf h}_{i}, \ i\neq k.\label{eq: Wbb-fi}
\end{align}

Then, we can rewrite the gradient descent iteration equation for updating digital precoder as
\begin{equation} \label{eq: gradient descent iteration-Wbb}
	{\bf w}_{\text{D}k}^{(l)}={\bf w}_{\text{D}k}^{(l-1)}-\delta\sum_{i=1}^K \rho_{k,i}^{\text{D}}(\hat{\bf h}_i,{\bf W}_{\text{D}}^{(l-1)})\hat{\bf h}_{i},
\end{equation}
where the function $\rho_{k,i}^{\text{D}}(\hat{\bf h}_i,{\bf W}_{\text{D}}^{(l-1)})$ is defined in \eqref{eq: Wbb-fk}~and~\eqref{eq: Wbb-fi}.

For analog precoder, to derive its gradient, we rewrite the term ${\bf h}_{k}^{\rm H} {\bf W}_{\text{A}} {\bf w}_{\text{D}i}$ as \(({\bf w}_{\text{D}i}^{\top} \otimes {\bf h}_k^H) \vec{\bf W}_{\text{A}}\), which is further denoted as $\bar{\bf h}_{k,i}\vec{\bf W}_{\text{A}}$ with $\bar{\bf h}_{k,i}\triangleq({\bf w}_{\text{D}i}^{\top} \otimes {\bf h}_k^H)^H \in {\mathbb C}^{N_s N \times 1}$ for notational simplicity. Then, the gradient of the objective function with respect to the vectorized analog precoder $\vec{\bf W}_{\text{A}}$  can be obtained as
\begin{equation} \label{eq: gradient-Wrf}
	\begin{aligned}
		\bigtriangledown_{\vec{\bf W}_{\text{A}}}&r
    =\sum_{k=1}^K \frac{\partial r(\gamma_1,\cdots,\gamma_K)}{\partial \gamma_{k}} \cdot \frac{\partial \gamma_{k}}{\partial \vec{\bf W}_{\text{A}}} \\
    =&\sum_{k=1}^K \!\Big(\underbrace{\Delta_{k,i}(\bar{\bf h}_{k,i},\vec{\bf W}_{\text{A}}^{(l-1)})}_{i=k} \!+ \!\sum_{i=1,i\neq k}^K \underbrace{\overline{\Delta}_{k,i}(\bar{\bf h}_{k,i},\vec{\bf W}_{\text{A}}^{(l-1)})}_{i\neq k}\Big).
	\end{aligned}
\end{equation}
Denoting $\frac {\partial r(\gamma_1,\cdots,\gamma_K)}{\partial \gamma_{k}}$ as $\lambda_{k}$, the terms $\Delta_{k,i}(\bar{\bf h}_{k,i},\vec{\bf W}_{\text{A}}^{(l-1)})$ and $\overline{\Delta}_{k,i}(\bar{\bf h}_{k,i},\vec{\bf W}_{\text{A}}^{(l-1)})$ can be obtained after some manipulations~as
\begin{align}
     \Delta_{k,i}(\bar{\bf h}_{k,i},\!\vec{\bf W}_{\text{A}}^{(l-1)}\!)&
     \!=\!\lambda_{k}\frac{\bar{\bf h}_{k,i}^{\rm H}\vec{\bf W}_{\text{A}}^{(l-1)}}{\sum_{j=1, j\neq k}^K |\bar{\bf h}_{k,j}^{\rm H} \vec{\bf W}_{\text{A}}^{(l-1)}|^2 \!+\! \sigma^2}\bar{\bf h}_{k,i}\nonumber\\
    &\triangleq \rho_{k,i}^{\text{A}}(\bar{\bf h}_{k,i},\vec{\bf W}_{\text{A}}^{(l-1)}) \bar{\bf h}_{k,i}, \ i= k.\label{eq: Wrf-fk}
\end{align}
\begin{align}
    \overline{\Delta}_{k,i}\!(\bar{\bf h}_{k,i},\!\vec{\bf W}_{\text{A}}^{(l-1)}\!)&
    \!=\!\lambda_{k}\frac{-|\bar{\bf h}_{k,k}^{\rm H}\vec{\bf W}_{\text{A}}^{(l-1)}|^2 \!\cdot\! \bar{\bf h}_{k,i}^{\rm H}\vec{\bf W}_{\text{A}}^{(l-1)}}{\left(\sum_{j=1,j\neq k}^K |\bar{\bf h}_{k,j}^{\rm H} \vec{\bf W}_{\text{A}}^{(l-1)}|^2 \!+\! \sigma^2\right)^2} \bar{\bf h}_{k,i} \nonumber\\
    &\triangleq \rho_{k,i}^{\text{A}}(\bar{\bf h}_{k,i},\vec{\bf W}_{\text{A}}^{(l-1)}) \bar{\bf h}_{k,i}, \ i\neq k.\label{eq: Wrf-fi}
\end{align}
Finally, the gradient descent iteration equation for updating analog precoder can be obtained as
\begin{equation} \label{eq: gradient descent iteration-Wrf}
	\vec{\bf W}_{\text{A}}^{(l)}=\vec{\bf W}_{\text{A}}^{(l-1)}-\delta \sum_{k=1}^K\sum_{i=1}^K \rho_{k,i}^{\text{A}}(\bar{\bf h}_{k,i},\vec{\bf W}_{\text{A}}^{(l-1)})\bar{\bf h}_{k,i},
\end{equation}
where the function $\rho_{k,i}^{\text{A}}(\bar{\bf h}_{k,i},\vec{\bf W}_{\text{A}}^{(l-1)})$ is defined in \eqref{eq: Wrf-fk} and~\eqref{eq: Wrf-fi}.

\subsection{Design of Gradient-Driven GNN}
By examining the iteration equations for digital and analog precoders given by \eqref{eq: gradient descent iteration-Wbb} and \eqref{eq: gradient descent iteration-Wrf}, we can find an interdependence: the gradient for digital precoder is affected by the analog precoded channel $\hat{\bf h}_i$ as defined above \eqref{eq: gradient-Wbb}, while the gradient for analog precoder is affected by the digital precoder involved in $\bar{\bf h}_{k,i}$  as defined above \eqref{eq: gradient-Wrf}. This interrelationship suggests an interleaved structure of the gradient-driven GNNs for learning hybrid precoders.

\begin{figure}
	\centering
	\includegraphics[width=0.48\textwidth]{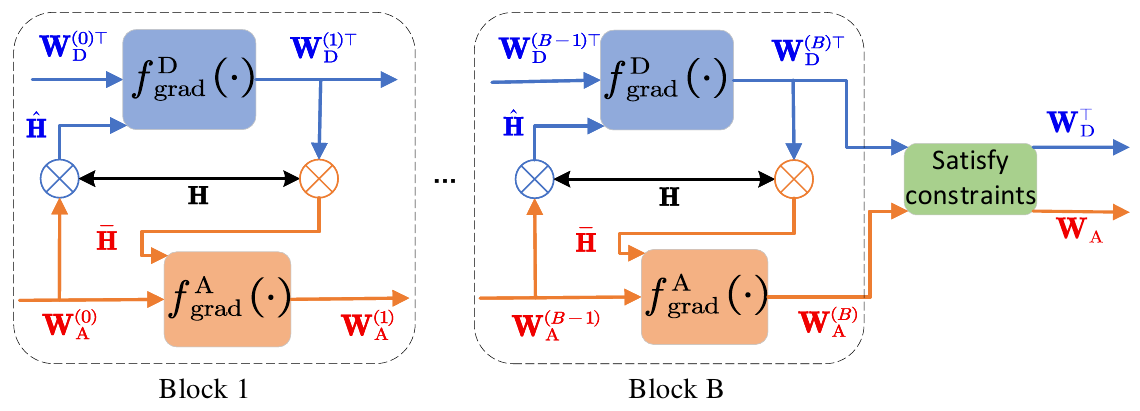}
	\caption{The structure of gradient-driven GNN for learning hybrid precoding.}
	\label{fig:hybrid-precoding}
\end{figure}

Figure \ref{fig:hybrid-precoding} shows the designed gradient-driven GNN. The input is channel matrix ${\bf H}$, and the outputs are digital precoder ${\bf W}_{\text{D}}$ and analog precoder ${\bf W}_{\text{A}}$. The GNN consists of hidden blocks, each containing two cascaded sub-networks, which are dedicated to updating the digital precoder (i.e., $f_{\text{grad}}^{\text{D}}(\cdot)$) and analog precoder (i.e., $f_{\text{grad}}^{\text{A}}(\cdot)$), respectively. 
The update equations for the two sub-networks in the $b$-th hidden block are denoted by
\begin{equation} \label{eq: update equation-Wbb}
	{\bf W}_{\text{D}}^{(b)}=f_{\text{grad}}^{\text{D}}({\bf W}_{\text{D}}^{(b-1)},\hat{\bf H}),
\end{equation}
\begin{equation} \label{eq: update equation-Wrf}
	\vec{\bf W}_{\text{A}}^{(b)}=f_{\text{grad}}^{\text{A}}(\vec{\bf W}_{\text{A}}^{(b-1)},\bar{\bf H}),
\end{equation}
where $\hat{\bf H} = [\hat{\bf h}_1, \cdots, \hat{\bf h}_K]\in {\mathbb C}^{N_s \times K}$, and $\bar{\bf H} = [\bar{\bf h}_{1,1}, \cdots, \bar{\bf h}_{1,K}, \bar{\bf h}_{2,1}, \cdots, \bar{\bf h}_{K,K}]\in {\mathbb C}^{N_s N \times KK}$

The sub-networks $f_{\text{grad}}^{\text{D}}(\cdot)$ and $f_{\text{grad}}^{\text{A}}(\cdot)$ need to satisfy the permutation properties defined in \eqref{eq: hybrid-PE}, i.e.,
\begin{equation} \label{eq: update equation-Wbb-PE}
	{\bf\Pi}_{\text{R}}^T{\bf W}_{\text{D}}^{(b)}{\bf\Pi}_{\text{U}}=f_{\text{grad}}^{\text{D}}({\bf\Pi}_{\text{R}}^T{\bf W}_{\text{D}}^{(b-1)}{\bf\Pi}_{\text{U}},{\bf\Pi}_{\text{R}}^T\hat{\bf H}{\bf\Pi}_{\text{U}}),
\end{equation}
\begin{align} \label{eq: update equation-Wrf-PE}
	&({\bf\Pi}_{\text{A}}^T\otimes{\bf\Pi}_{\text{R}}^T)\vec{\bf W}_{\text{A}}^{(b)} \nonumber\\
 &=f_{\text{grad}}^{\text{A}}\left(({\bf\Pi}_{\text{A}}^T\otimes{\bf\Pi}_{\text{R}}^T) \vec{\bf W}_{\text{A}}^{(b-1)},({\bf\Pi}_{\text{A}}^T\otimes{\bf\Pi}_{\text{R}}^T)\bar{\bf H}({\bf\Pi}_{\text{U}}\otimes{\bf\Pi}_{\text{U}})\right).
\end{align}
Specifically, the sub-network $f_{\text{grad}}^{\text{D}}(\cdot)$ for updating ${\bf W}_{\text{D}}$ is equivariant to the permutation of user and RF chain indices. The sub-network $f_{\text{grad}}^{\text{A}}(\cdot)$ for updating ${\bf W}_{\text{A}}$ is equivariant to the permutation of antenna, RF chain indices, and invariant to the permutation of user indices.

Given the three types of indices, i.e., users, RF chains, and antennas, we construct the graph for the two GNN sub-networks as follows. The graph consists of three types of vertices (i.e., $K$ user vertices, $N$ antenna vertices, and $N_s$ RF chain vertices), and two types of edges (i.e., edge $(k, m)$ that connects the $k$-th user vertex and $m$-th RF chain vertex, and edge $(n, m)$ that connects the $n$-th antenna vertex and $m$-th RF chain vertex). The design of the update equations for the two GNN sub-networks is detailed below.

\subsubsection{\underline{ Sub-network $f_{\text{grad}}^{\text{D}}(\cdot)$}}
Similar to the design of the gradient-driven GNN for fully digital precoding, inspired by the form of the gradient descent iteration equation in \eqref{eq: gradient descent iteration-Wbb} for digital precoder, we design the update equation of the sub-network as
\begin{equation} \label{eq: combination equation-Wbb}
    {d}_{k,m,t}^{\text{D}(l)}=\sigma\left(s^{(l)}{d}_{k,m,t}^{\text{D}(l-1)}+p^{(l)}\sum_{i=1}^K \alpha_{k,i,t}^{\text{D}(l-1)} \hat{h}_{i,m}\right),
\end{equation}
where only the $t$-th hidden representation is considered for simplicity, which can easily be extended to multiple hidden representations, as done in Sec.~\ref{extension-digital}.

In \eqref{eq: combination equation-Wbb}, $s^{(l)}$, $p^{(l)}$, and $\alpha_{k,i,t}^{\text{D}(l-1)}$ are learnable parameters similar to those in \eqref{eq: combination equation-V}. The attention coefficient $\alpha_{k,i,t}^{\text{D}(l-1)}$ reflects the importance of the information provided by user \(i\) to user \(k\). By comparing \eqref{eq: Wbb-fk} and \eqref{eq: Wbb-fi} with \eqref{eq: fk} and \eqref{eq: fi}, we observe a similar expression, except that the term $\mathbf{h}_i$ in the latter is replaced by $\hat{\bf h}_i$. Thus, $\alpha_{k,i,t}^{\text{D}(l-1)}$ can still be learned by the network designed in \eqref{eq: attention-NN-V}, which can be expressed as
\begin{align} \label{eq: attention-NN-Wbb}
	\alpha_{k,i,t}^{\text{D}(l-1)}=&\phi\Big(\hat{\bf h}_{i}^{\rm H}\bar{\bf d}_{k,t}^{\text{D}(l-1)},\ \textstyle\sum_{j=1, j\neq i}^K |\hat{\bf h}_{i}^{\rm H}\bar{\bf d}_{j,t}^{\text{D}(l-1)}|^2, \nonumber\\
&\qquad |\hat{\bf h}_{i}^{\rm H}\bar{\bf d}_{i,t}^{\text{D}(l-1)}|^2\cdot\xi_{i,k}\Big).
\end{align}

\subsubsection{\underline{Sub-network $f_{\text{grad}}^{\text{A}}(\cdot)$}}
Based on the gradient descent iteration equation in \eqref{eq: gradient descent iteration-Wrf}, the update equation for the analog precoder is designed as
\begin{equation} \label{eq: combination equation-Wrf}
    \Vec{\bf D}_{t}^{\text{A}(l)}=\sigma\left(s^{(l)}\Vec{\bf D}_{t}^{\text{A}(l-1)}+p^{(l)}\sum_{k=1}^K\sum_{i=1}^K \alpha_{k,i,t}^{\text{A}(l-1)} \bar{\bf h}_{k,i}\right),
\end{equation}
where $\Vec{\bf D}_{t}^{\text{A}(l)}\in {\mathbb C}^{N_s N \times 1}$ is the $t$-th hidden representation of the $l$-th layer. \eqref{eq: combination equation-Wrf} can be extended to multiple hidden representations as done in Sec.~\ref{extension-digital}.

%

For the learning of the attention coefficient $\alpha_{k,i,t}^{\text{A}(l-1)}$, 
by comparing the expressions of $\rho_{k,i}^{\text{A}}(\bar{\bf h}_{k,i},{\bf W}_{\text{A}}^{(l-1)})$ in \eqref{eq: Wrf-fk} and \eqref{eq: Wrf-fi} with the expressions of $\rho_{k,i}({\bf h}_i,{\bf V}^{(l-1)})$ in \eqref{eq: fk} and \eqref{eq: fi}, we can observe a clear similarity between them. Then, by replacing the input ${\bf h}_i$ and $\bar{\bf{d}}_{k,t}$ of $\phi(\cdot)$ in \eqref{eq: attention-NN-V} to $\bar{\bf h}_{k,i}$ and $\vec{\bf D}_{t}^{\text{A}}$, respectively, we obtain the attention mechanism~as
\begin{align} \label{eq: attention-NN-Wrf}
	\alpha_{k,i,t}^{\text{A}(l-1)}=&\phi\Big(\bar{\bf h}_{k,i}^{\rm H}\Vec{\bf D}_{t}^{\text{A}(l-1)},\ \textstyle\sum_{j=1, j\neq k}^K |\bar{\bf h}_{k,j}^{\rm H}\Vec{\bf D}_{t}^{\text{A}(l-1)}|^2, \nonumber\\
&\qquad |\bar{\bf h}_{k,k}^{\rm H}\Vec{\bf D}_{t}^{\text{A}(l-1)}|^2\cdot\xi_{i,k}\Big).
\end{align}

The two designed sub-networks satisfy the permutation equivariant and invariant properties as given in \eqref{eq: update equation-Wbb-PE} and \eqref{eq: update equation-Wrf-PE}. This can be verified by using the same approach in Sec. \ref{PE-digital}, which are not detailed here to avoid redundancy.

\subsection{Exemplary Problem: SE Maximization}
We take the SE maximization problem as an example to demonstrate the application of the proposed GNN.

The hybrid precoder optimization problem that maximizes the sum SE subject to the power constraint of the BS and the constant modulus constraint of analog precoder can be expressed~as
\begin{subequations}\label{P4: max sum-rate-hybrid-SE}
	\begin{align}
		\max_{{\bf W}_{\text{A}},{\bf W}_{\text{D}}} ~~& \sum_{k=1}^{K} \log \left(1 + \gamma_{k}\right) \label{P4: msr-1} \\
		{\rm s.t.}\  ~~& |({\bf W}_{\text{A}})_{j,l}|=1,j=1,\cdots,N,l=1,\cdots,N_s, \label{P4: msr-2} \\
        ~~& \|{\bf W}_{\text{A}}{\bf W}_{\text{D}}\|_F^2 \leq P^{\max}.
        \label{P4: msr-3}
	\end{align}
\end{subequations}
Employing the proposed gradient-driven GNN to solve problem \eqref{P4: max sum-rate-hybrid-SE} is straightforward. First, we employ the negative sum SE as the loss function during training. Second, the output block uses $({\bf W}_{\text{A}})_{j,l}=({\bf W}_{\text{A}}^{(B)})_{j,l}/|({\bf W}_{\text{A}}^{(B)})_{j,l}|$ to ensure the constant modulus constraint in \eqref{P4: msr-2}, and uses ${\bf W}_{\text{D}}={\bf W}_{\text{D}}^{(B)}/\|{\bf W}_{\text{A}}{\bf W}_{\text{D}}^{(B)}\|_F$ to ensure the power constraint in \eqref{P4: msr-3}, where $B$ is the total number of hidden blocks.

The proposed gradient-GNN is also applicable to learn hybrid precoding policies for problems with other objective functions related to SINR, e.g., Log-SE maximization, which are not detailed here due to lack of space.

\section{Simulation Results} \label{simulation}
In this section, we evaluate the performance of the proposed gradient-driven GNN for learning different precoding policies with different objective functions. The training complexity and inference time against various baseline numerical and learning methods are also compared.

\subsection{Learning Fully Digital Precoding for SE Maximization}
In the simulations of fully digital precoding for SE maximization, we generate 100,000 channel samples for training the GNNs, and 1,000 channel samples for testing. The channel samples are drawn from the Rayleigh channel model. For evaluating the generalization performance, different numbers of antennas and users are considered. The SNR is set to 10~${\rm dB}$. For the proposed GNN, we use Tanh as the activation function and Adam as the optimizer, and set the batch size as 200 and learning rate as 0.001. The GNN includes four hidden layers each with 128 hidden edge representations, and the employed attention mechanism is implemented using a single-layer fully-connected neural network. We use unsupervised learning to train the network, where the loss function is the negative SE. For the compared baseline GNNs, we have fine-tuned their hyper-parameters. All the results are obtained on a personal computer with Intel CoreTM i9-9920X CPU and Nvidia RTX 2080Ti~GPU.

\begin{itemize}
    \item \textbf{WMMSE}: This is a widely used numerical algorithm for optimizing SE-maximal digital precoding \cite{WMMSE}, which can find at least a local optimal precoder.
        \item \textbf{Pinv-GNN}: This is the GNN proposed in \cite{catDHD}, which was designed by using the knowledge of Taylor's expansion of matrix pseudo-inverse. During training, the number of users $K$ is fixed to $6$ or $12$ and the number of antennas $N$ is fixed to $12$.
    \item \textbf{Pinv-GNN-exp}: This is also the {Pinv-GNN} method, but it employs a different dataset generation method to improve the generalization performance of {Pinv-GNN}~\cite{catDHD}. Specifically, to evaluate the generalization performance concerning the number of users, the training dataset includes a variable number of users $K$, which is randomly sampled from an exponential distribution with the mean of 2 and the standard deviation of 2. The values of $K$ are truncated at eight, i.e., all samples are with $K\leq 8$. To evaluate the generalization performance related to the number of antennas, the training dataset comprises a variable number of antennas $N$, which is also randomly drawn from an exponential distribution with the mean of 12 and the standard deviation of 10. All training samples are with $N\leq 15$.
    \item \textbf{Vanilla-GNN}: This is the vanilla-GNN designed in \cite{ZBC}.
    \item \textbf{GCN-WMMSE}: This is the deep unfolding network proposed in \cite{WMMSE4}, where the structure of GCN is applied to unfold the WMMSE algorithm \cite{WMMSE}.
    \item \textbf{GAT}: This is the GAT designed in \cite{GAT}, which is based on a homogeneous GNN. The employed graph consists solely of user vertices, with the channel vector of each user serving as the vertex feature. An attention mechanism is applied to update the hidden representation of the~vertices.    
    \item \textbf{Gradient-GNN}: This is the proposed GNN. During training, the number of users $K$ is fixed to $6$ or $12$ and the number of antennas $N$ is fixed to $12$.
    \item \textbf{Gradient-GNN-exp}: This is also the proposed GNN, which uses the same training dataset as Pinv-GNN-exp.
\end{itemize}

\subsubsection{\underline{Performance Comparison}}
We compare the performance of the following methods.
\begin{figure}
	\centering
	\vspace{-3mm}
	\includegraphics[width=0.45\textwidth]{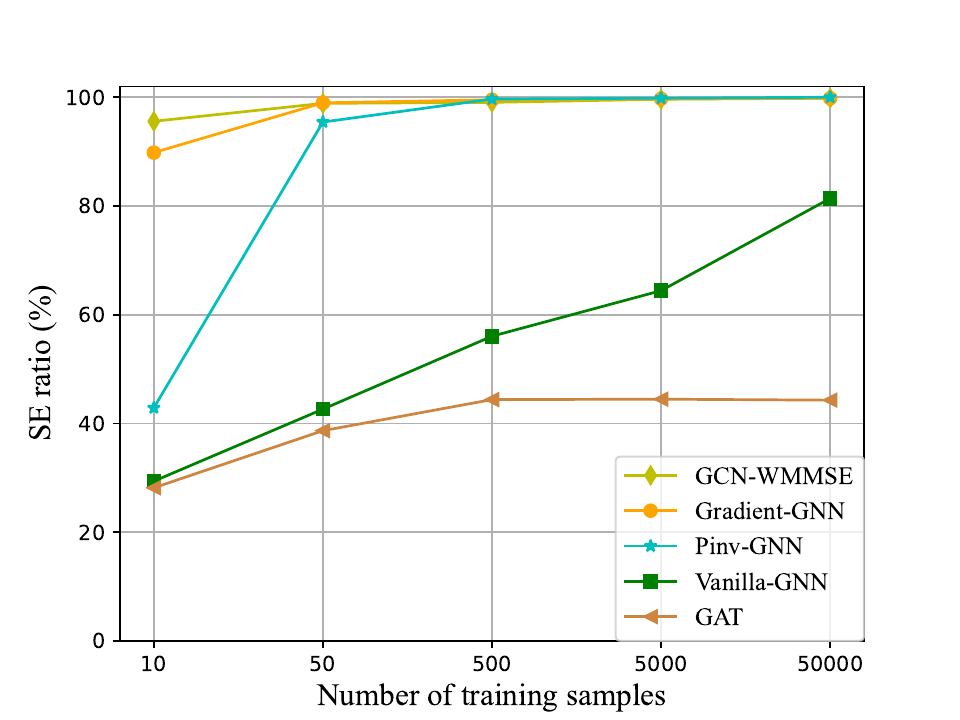}
	\caption{Learning performance vs the number of training samples, $K=6$ and $N=12$.}
	\label{fig:sample-K6}
\end{figure}

\begin{figure}
	\centering
	\vspace{-3mm}
	\includegraphics[width=0.45\textwidth]{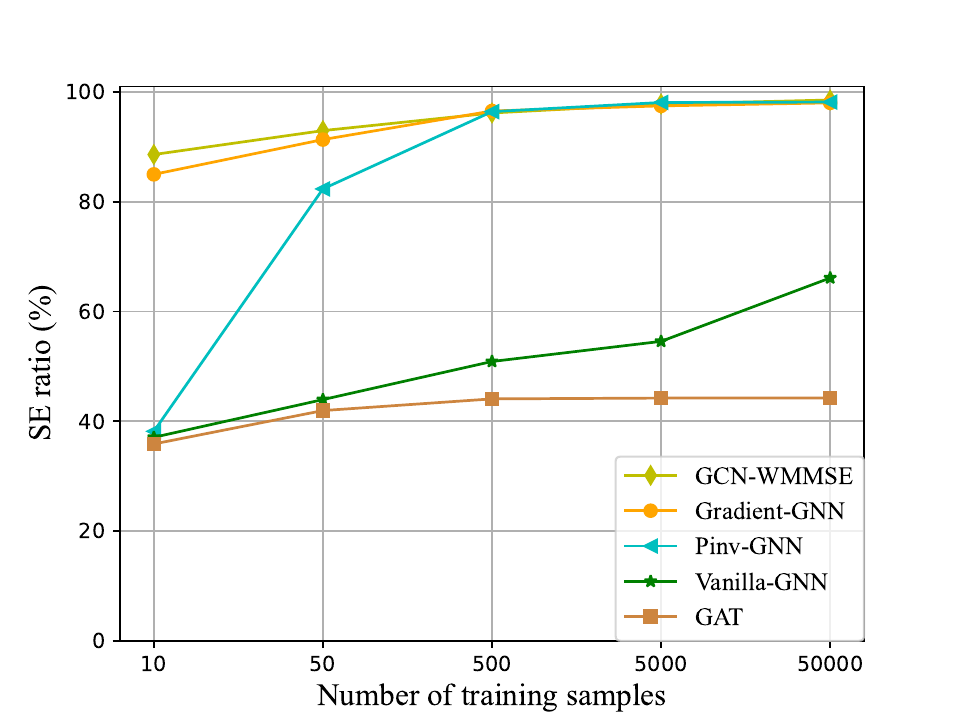}
	\caption{Learning performance vs the number of training samples, $K=12$ and $N=12$.}
	\label{fig:sample-K12}
	\vspace{-1mm}
\end{figure}

In Fig. \ref{fig:sample-K6} and Fig. \ref{fig:sample-K12}, the learning performance with different numbers of training samples is depicted, where the GNNs are trained with $K=6$ and $K=12$ users, respectively. The Y-axis is the SE ratio achieved by GNNs to that of WMMSE. It can be seen that the proposed GNN can achieve 99.0\% ($K=6$) and 91.3\% ($K=12$) performance of WMMSE with only 50 training samples. With the increase of training samples, the gap between the proposed GNN and Pinv-GNN decreases, but the gains over Vanilla-GNN and GAT are evident. The GCN-WMMSE achieves a slightly higher SE ratio than the proposed GNN with fewer training samples. However, the GCN-WMMSE cannot be used to learn other objective functions and other precoding policies, because the unfolded WMMSE algorithm is tailored to the fully digital precoding for SE maximization.
\begin{figure}
	\centering
	\vspace{-3mm}
	\includegraphics[width=0.45\textwidth]{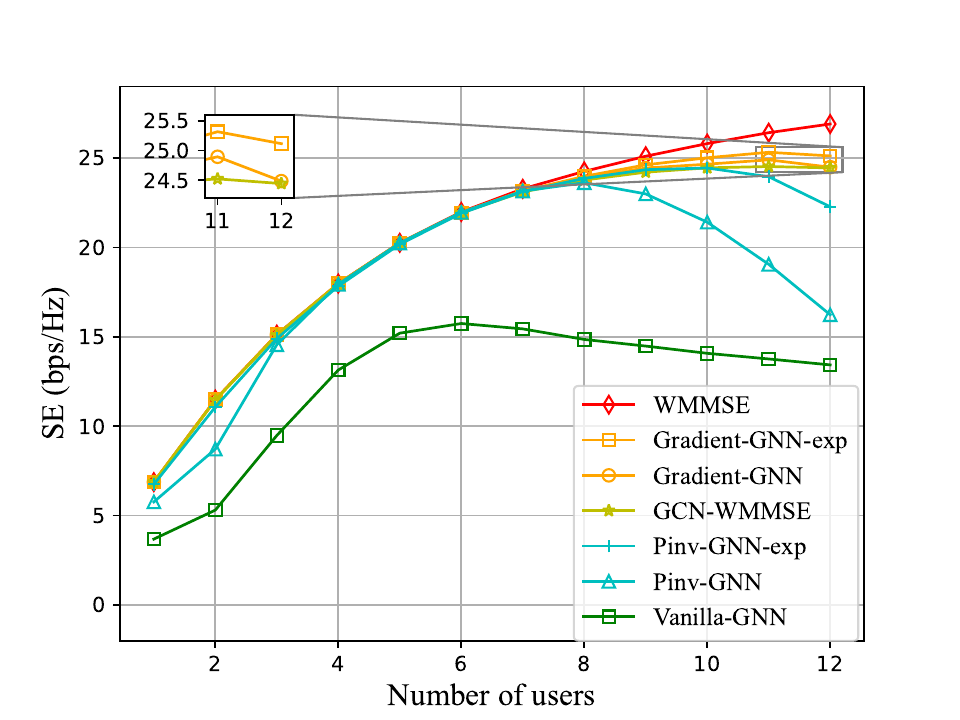}
	\vspace{-2mm}
	\caption{Generalization performance to the number of users with $N=12$ (Gradient-GNN-exp and Pinv-GNN-exp are trained with exponentially distributed $K$ while others are trained with $K=6$).}
	\label{fig:K6N12-userG}
	\vspace{-2mm}
\end{figure}

\begin{figure}
	\centering
	\vspace{-3mm}
	\includegraphics[width=0.45\textwidth]{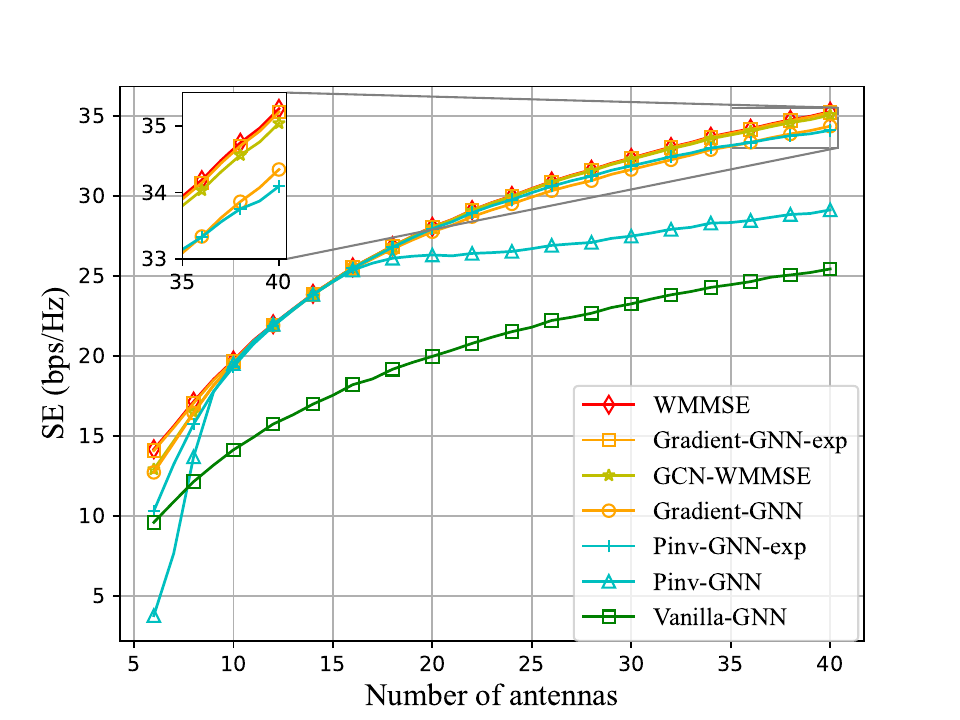}
	\vspace{-2mm}
	\caption{Generalization performance to the number of antennas with $K=6$ (Gradient-GNN-exp and Pinv-GNN-exp are trained with exponentially distributed $N$ while others are trained with $N=12$).}
	\label{fig:K6N12-anteG}
	\vspace{-3mm}
\end{figure}

In Fig. \ref{fig:K6N12-userG}, the generalization performance to the number of users is evaluated, where the number of antennas $N$ is 12. In the training phase, the number of users is set to $K=6$ for Gradient-/Pinv-/Vanilla-GNN/GCN-WMMSE. In the inference phase, we set the numbers of users ranging from $1$ to $12$. It can be seen that the proposed GNN performs close to WMMSE for small and immediate $K$, slightly exceeds GCN-WMMSE for large $K$, and outperforms Pinv-GNN for small and large $K$. The vanilla-GNN achieves poor generalization performance, especially when the numbers of users for training and testing have large difference. When the number of users follows an exponential distribution during training, the generalization performance of both the proposed GNN and Pinv-GNN improves for large $K$, and the gain of the proposed GNN is still evident for large $K$.

In Fig. \ref{fig:K6N12-anteG}, the generalization performance to the number of antennas is evaluated, where the number of users $K$ is 6. During the training phase, the number of antennas is set to $N=12$ for Gradient-/Pinv-/Vanilla-GNN/GCN-WMMSE, while during the inference phase, the number of antennas ranges from $6$ to $40$. It is shown that the proposed GNN has a significant gain over Pinv-GNN and Vanilla-GNN. The SE of GCN-WMMSE is slightly higher than the proposed GNN when generalizing to large $N$. When the number of antennas follows an exponential distribution during training, the generalization performance of both the proposed GNN and Pinv-GNN improves for large and small $N$, and the gain of the proposed GNN is still evident for small~$N$.

\begin{figure}
	\centering
	\vspace{-3mm}
	\includegraphics[width=0.45\textwidth]{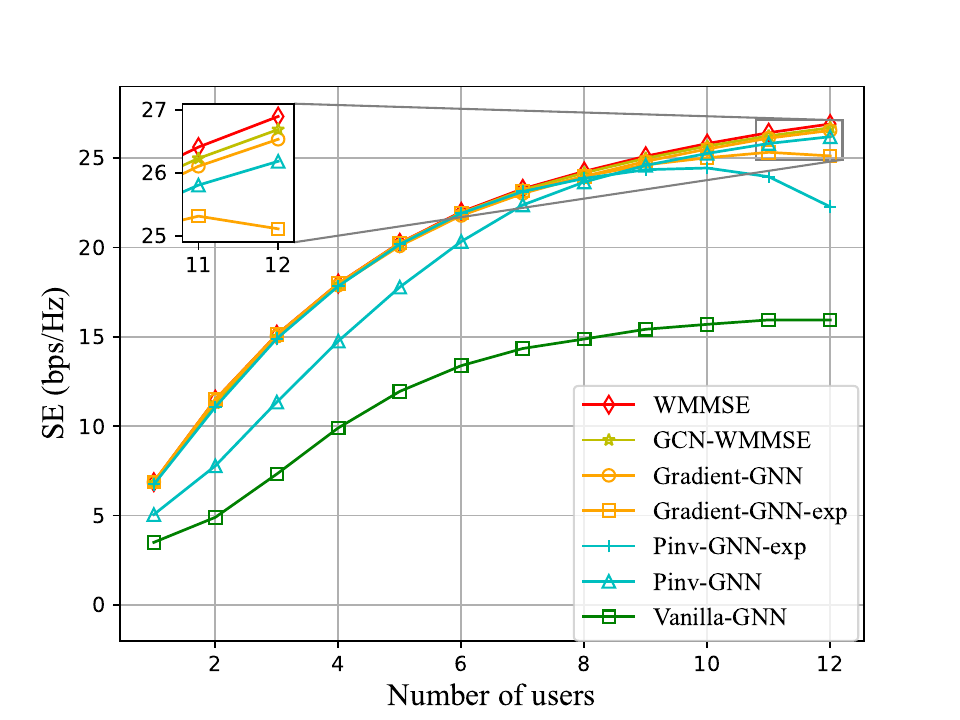}
	\caption{Generalization performance to the number of users with $N=12$ (Gradient-GNN-exp and Pinv-GNN-exp are trained with exponentially distributed $K$ while others are trained with $K=12$).}
	\label{fig:K12N12-userG}
	\vspace{-3mm}
\end{figure}

\begin{figure}
	\centering
	\vspace{-4mm}
	\includegraphics[width=0.45\textwidth]{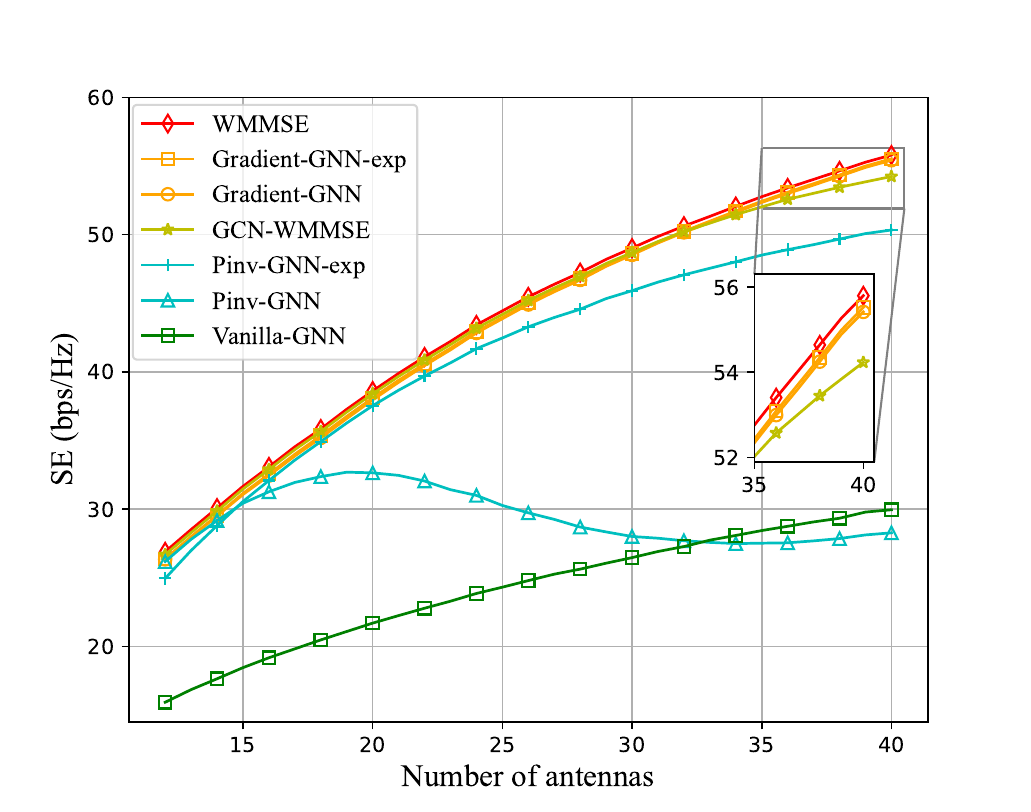}
	\vspace{-2mm}
	\caption{Generalization performance to the number of antennas with $K=12$ (Gradient-GNN-exp and Pinv-GNN-exp are trained with exponentially distributed $N$ while others are trained with $N=12$).}
	\label{fig:K12N12-anteG}
	\vspace{-3mm}
\end{figure}

In Fig. \ref{fig:K12N12-userG} and Fig. \ref{fig:K12N12-anteG}, we change the number of users to $K=12$ for Gradient-/Pinv-/Vanilla-GNN/GCN-WMMSE, and evaluate the generalization performance to the number of users and to the number of antennas, respectively. We can observe similar results to those in Fig.~\ref{fig:K6N12-userG} and Fig. \ref{fig:K6N12-anteG}, where the proposed GNN demonstrates good generalization performance.

\subsubsection{\underline{Complexity Comparison}}
We compare the complexity of the considered GNNs, where $K=6, 12$ and $N=12$. Training complexity includes space complexity, sample complexity, and time complexity. Specifically, space complexity and sample complexity are defined as the required minimal number of training parameters and samples for achieving an expected performance (set to 99\% of the performance of WMMSE for $K=6$ and 95\% for $K=12$). Time complexity is the training time with the minimal number of training parameters and samples to achieve the expected performance. Inference time is the runtime averaged over the 1000 test samples in the inference stage. The results are listed in Table~\ref{table: cplxty1} and Table~\ref{table: cplxty2}.

It is shown that compared with Pinv-GNN, the proposed GNN has lower space complexity and sample complexity with an increase of time complexity and inference time. GCN-WMMSE has lower space complexity, slightly higher sample complexity, and much higher training time and inference time than the proposed GNN. The Vanilla-GNN has very high complexity, which only achieves 81.4\% ($K=6$) and 66.1\% ($K=12$) performance of WMMSE even if trained with 100,000 samples.

\begin{table}[htb!]
	\centering
	\vspace{-2mm}
	\caption{Training complexity and inference time with $K=6$.}\label{table: cplxty1}
	\vspace{-2mm}
	\footnotesize
	\renewcommand\arraystretch{1.1}
        \setlength\tabcolsep{3pt}
	\begin{tabular}{c|c|c|c|c}
		\hline
		~ & \makecell{Grad-GNN\\$(K=6)$} & \makecell{Pinv-GNN\\$(K=6)$} &  \makecell{GCN-WMMSE\\$(K=6)$} & \makecell{Vanilla-GNN\\$(K=6)$}\\
		\hline
		Space & 132,098 & 299,008 & 67,604 & $>$3,156,992   \\
		\hline
		Sample & 100 & 200 & 200 & $>$100,000 \\
		\hline
		\makecell{Training\\(GPU)} & 85.58 s & 40.57 s & 938.15 s & $>$4519 s   \\
		\hline
		\makecell{Inference\\(GPU)} & 5.76 ms & 4.98 ms & 109.70 ms & $>$2.99 ms  \\
		\hline
		\makecell{Inference\\(CPU)} & 6.93 ms & 6.54 ms & 59.83 ms & $>$3.98 ms    \\
		\hline
	\end{tabular}
	\vspace{-2mm}
\end{table}

\begin{table}[htb!]
	\centering
	\vspace{-2mm}
	\caption{Training complexity and inference time with $K=12$.}\label{table: cplxty2}
	\vspace{-2mm}
	\footnotesize
	\renewcommand\arraystretch{1.1}
        \setlength\tabcolsep{3pt}
	\begin{tabular}{c|c|c|c|c}
		\hline
		~ & \makecell{Grad-GNN\\$(K=12)$} & \makecell{Pinv-GNN\\$(K=12)$} & \makecell{GCN-WMMSE\\$(K=12)$} & \makecell{Vanilla-GNN\\$(K=12)$}  \\
		\hline
		Space & 132,098 & 299,008 & 67,604 & $>$ 3,156,992  \\
		\hline
		Sample & 200 & 400 & 300 & $>$100,000\\
		\hline
		\makecell{Training\\(GPU)} & 92.72 s & 77.40 s & 2239.75 s& $>$7392 s   \\
		\hline
		\makecell{Inference\\(GPU)} & 5.99 ms & 5.51 ms & 217.41 ms & $>$3.49 ms \\
		\hline
		\makecell{Inference\\(CPU)} & 7.45 ms & 6.98 ms & 88.76 ms & $>$4.98 ms \\
		\hline
	\end{tabular}
	\vspace{-2mm}
\end{table}

\subsection{Learning Fully Digital Precoding for Log-SE Maximization}
For the log-SE maximization problem in \eqref{P2: max log-sum-rate}, the original WMMSE algorithm is no longer applicable. In~\cite{WMMSE-PF}, a numerical algorithm known as WMMSE-PF was proposed to solve problem \eqref{P2: max log-sum-rate}, which is a revised version of the WMMSE algorithm and thus is used as the performance baseline in the simulations. The simulation setup is the same as the previous subsection.

\begin{figure}
	\centering
	\vspace{-3mm}
	\includegraphics[width=0.45\textwidth]{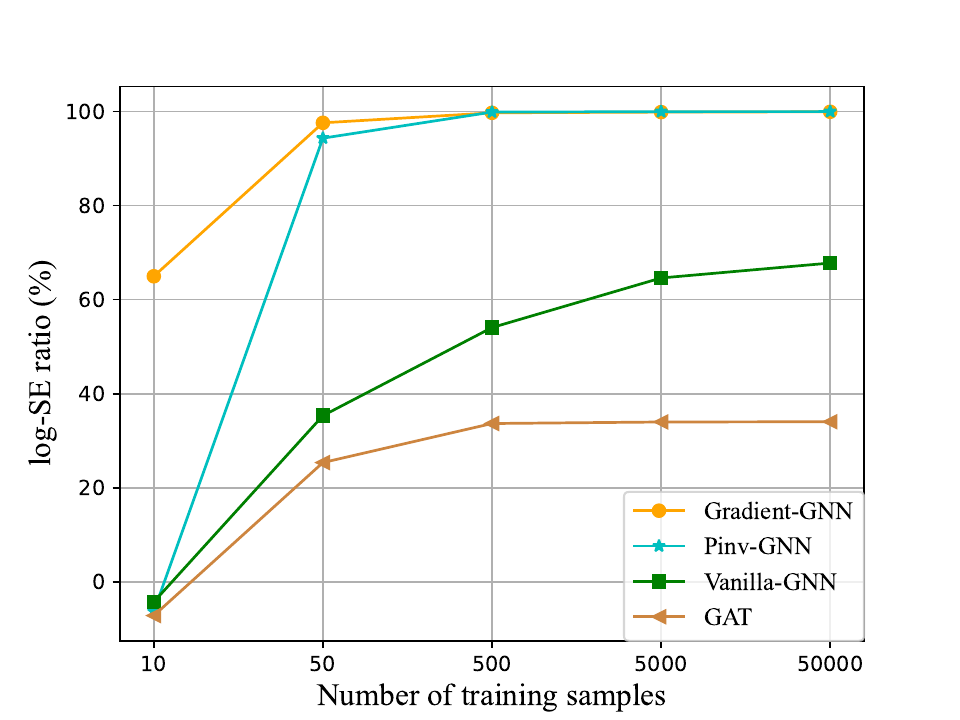}
	\caption{Learning performance vs the number of training samples, $K=6$ and $N=12$.}
	\label{fig:log-K6-sample}
	\vspace{-2mm}
\end{figure}

In Fig. \ref{fig:log-K6-sample}, we evaluate the learning performance with varying numbers of training samples, where the GNNs are trained with $K=6$ and $N=12$. The Y-axis denotes the ratio of the logarithmic SE achieved by the GNNs to those achieved by WMMSE-PF. It can be observed that the proposed GNN attains 97.5\% of the WMMSE-PF performance with merely 50 training samples. As the number of training samples increases, the performance gap between the proposed GNN and Pinv-GNN diminishes, and the improvements over Vanilla-GNN and GAT are significant.

\begin{figure}
	\centering
	\vspace{-3mm}
	\includegraphics[width=0.45\textwidth]{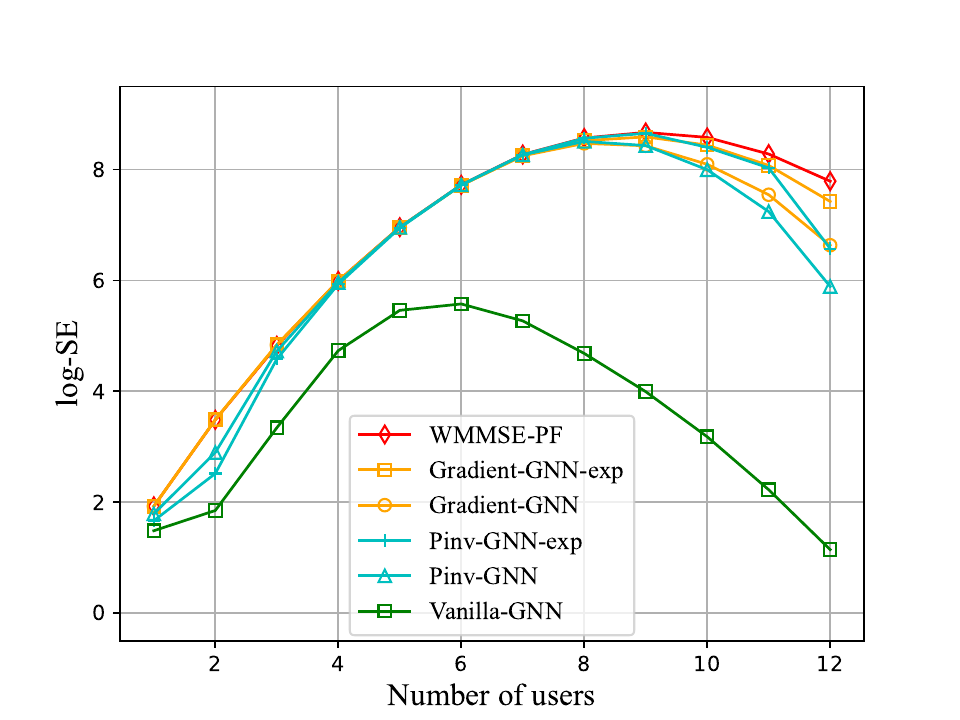}
	\caption{Generalization performance to the number of users  with $N=12$ (Gradient-GNN-exp and Pinv-GNN-exp are trained with exponentially distributed $K$ while others are trained with $K=6$).}
	\label{fig:log-userG-K6N12}
	\vspace{-2mm}
\end{figure}

In Fig. \ref{fig:log-userG-K6N12},  the generalization performance to the number of users is evaluated, where the number of antennas $N$ is 12. In the training phase, the number of users is set to $K=6$ during the training phase for Gradient-/Pinv-/Vanilla-GNN. During the inference phase, we evaluate the generalization performance by varying the number of users from 1 to 12. The results indicate that the proposed GNN surpasses the performance of Pinv-GNN. In contrast, Vanilla-GNN demonstrates poor performance, particularly when there is a significant discrepancy between the number of users in the training and testing phases. When the number of users follows an exponential distribution during training, the generalization performance of both the proposed GNN and Pinv-GNN improves for large $K$, and the advantages of the proposed GNN remain evident for small and large $K$.

\begin{figure}
	\centering
	\vspace{-3mm}
	\includegraphics[width=0.45\textwidth]{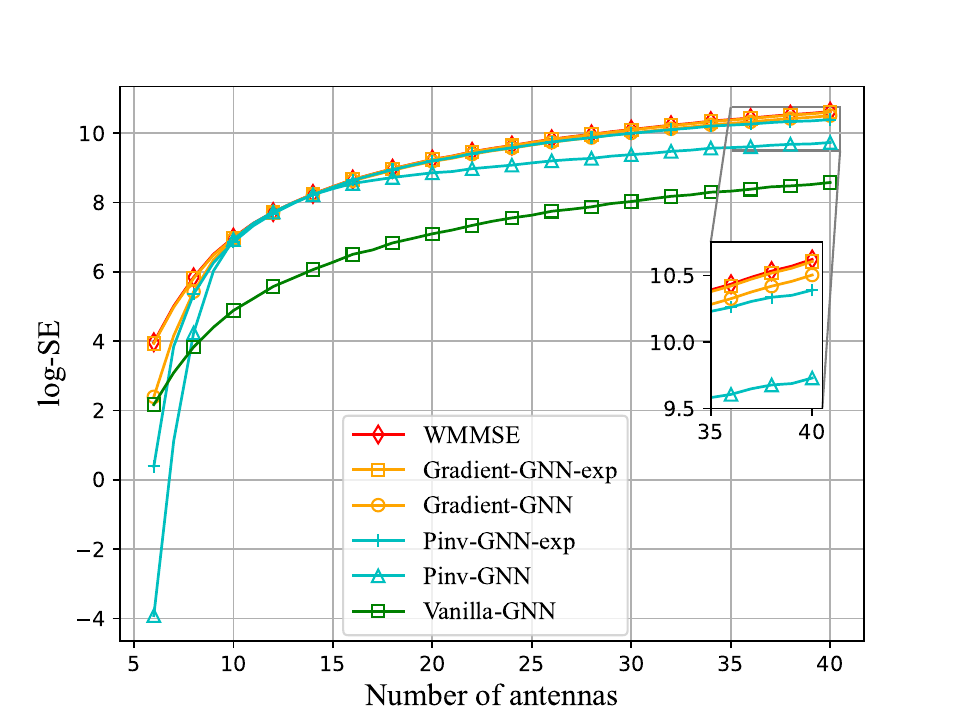}
	\caption{Generalization performance to the number of antennas  with $K=6$ (Gradient-GNN-exp and Pinv-GNN-exp are trained with exponentially distributed $N$ while others are trained with $N=12$).}
	\label{fig:log-antG-K6N12}
	\vspace{-3mm}
\end{figure}

In Fig. \ref{fig:log-antG-K6N12}, we evaluate the generalization performance with respect to the number of antennas, where $K$ is 6. The number of antennas $N$ is fixed to 12 for Gradient-/Pinv-/Vanilla-GNN during training and $N$ varies from 6 to 40 during testing. The results demonstrate that the proposed GNN exhibits a substantial performance gain over the other GNNs, particularly as the number of antennas increases. When the number of antennas follows an exponential distribution during training, the generalization performance of both the proposed GNN and Pinv-GNN improves, but for small~$N$, the proposed GNN continues to outperform the Pinv-GNN.

\subsection{Learning Hybrid Precoding}
In this subsection, we evaluate the performance and complexity of the proposed gradient-GNN for learning hybrid precoders, where the SE maximization problem in \eqref{P4: max sum-rate-hybrid-SE} is considered.

In the simulations, we generate 100,000 channel samples for training the GNNs and 1,000 channel samples for testing. To evaluate the generalization performance of the GNNs, we consider different values of \(K\) and \(N\). The channel samples are generated from the Saleh-Valenzuela (SV) channel model, which effectively captures the characteristics of mmWave channels. A uniform linear antenna array is considered, and the channel from the base station to a user is modeled as
\[{\bf h}=\sqrt{\frac{N}{N_{cl}N_{ray}}}\sum_{i=1}^{N_{cl}}\sum_{j=1}^{N_{ray}}\beta_{i,j}{\bf f}(\theta_{i,j})\in {\mathbb C}^{N},\]
where \(N_{cl}\) is the number of scattering clusters, \(N_{ray}\) is the number of scattering rays, \(\beta_{i,j}\) is a complex gain, and \({\bf f}(\theta_{i,j})\) is the array response. We set \(N_{cl}=4\) , \(N_{ray}=5\), and \(\beta_{i,j}\sim\mathcal{CN}(0,1)\).

The proposed GNN includes four hidden layers, each with 64 hidden edge representations. We use the Adam optimizer with the batch size of 200 and the learning rate of 0.001. The network is trained using an unsupervised learning approach, where the loss function is the negative SE.

\subsubsection{\underline{Performance Comparison}}
We compare the performance of the following methods.
\begin{itemize}
        \item \textbf{MO}: This is an alternating minimization algorithm based on manifold optimization to approach the performance of the fully digital precoder~\cite{MO}.
        \item \textbf{OMP}: This is the orthogonal matching pursuit algorithm to solve sparse reconstruction problem~\cite{OMP}.
        \item \textbf{RGNN}: This is a size-generalizable GNN proposed in \cite{RGNN}, which was designed in a recursive manner.
	\item \textbf{RGNN-exp}: This is also RGNN, but it uses a training dataset with different numbers of users $K$ to improve the generalization performance of RGNN, as done in \cite{RGNN}. Specifically, $K$ is randomly drawn during training from an exponential distribution with the mean of 3 and the standard deviation of 2. 
        \item \textbf{3D-GNN}: This is the 3D-GNN designed in \cite{MDGNN}.
        \item \textbf{3D-GNN-exp}: Similar to RGNN-exp, this method trains 3D-GNN using a dataset with exponentially distributed~$K$.
        \item \textbf{CHP-ULGAT}: This is a graph attention network based method for learning the hybrid precoder designed in \cite{GAT+FCNN}.
        \item \textbf{Gradient-GNN}: This is the proposed GNN designed for hybrid precoding.
        \item \textbf{Gradient-GNN-exp}: Similar to RGNN-exp, this method trains the proposed Gradient-GNN using a dataset with exponentially distributed $K$.
\end{itemize}

\begin{figure}
	\centering
	\vspace{-3mm}
	\includegraphics[width=0.45\textwidth]{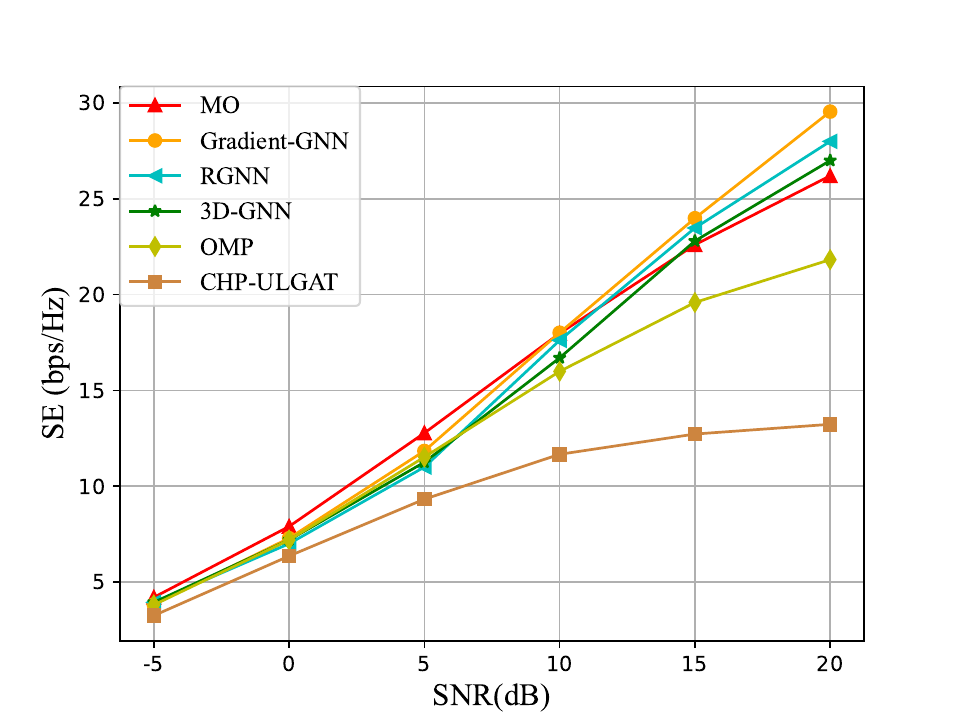}
	\vspace{-2mm}
	\caption{Learning performance vs different SNRs, $K=4$, $N_s=4$ and $N=16$.}
	\label{fig:SNR-16-4-4}
	\vspace{-3mm}
\end{figure}

Fig. \ref{fig:SNR-16-4-4} shows the learning performance with different SNRs, where the GNNs are trained with \(K=4\), \(N_s=4\), \(N=16\), and the Y-axis is the achieved SE. It is shown that Gradient-GNN, RGNN, and 3D-GNN are superior to OMP and CHP-ULGAT when SNR is high, and the proposed Gradient-GNN achieves slightly higher SE than RGNN and 3D-GNN.

\begin{figure}
	\centering
	\vspace{-3mm}
	\includegraphics[width=0.45\textwidth]{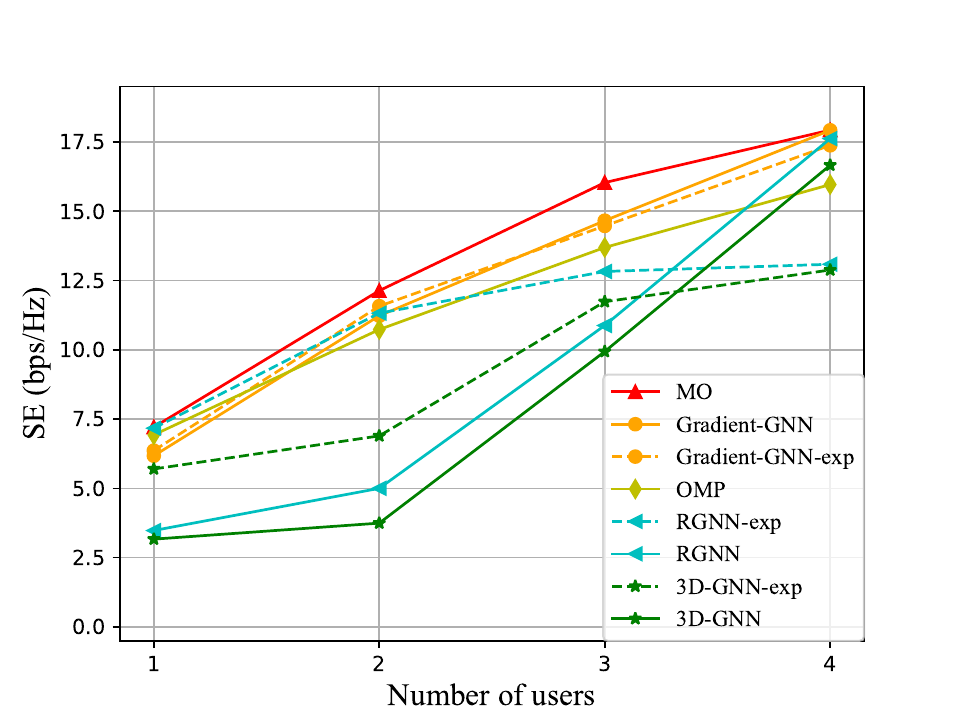}
	\vspace{-2mm}
	\caption{Generalization performance to the number of users with $N_s=4$ and $N=16$ (Gradient-GNN-exp, RGNN-exp, and 3D-GNN-exp are trained with exponentially distributed $K$ while others are trained with $K=4$).}
	\label{fig:userG-16-4-4}
	\vspace{-2mm}
\end{figure}

In Fig. \ref{fig:userG-16-4-4}, the generalization performance to the number of users is evaluated, where the number of RF chains $N_s$ is 4 and the number of antennas $N$ is 16. In the training phase, the number of users is set to $K=4$ for Gradient-GNN/RGNN/3D-GNN, and the SNR is set to 10 dB. During the inference phase, we evaluate the generalization performance by varying the number of users from 1 to 4. The results indicate that the proposed GNN performs close to MO, and surpasses RGNN and 3D-GNN. When the number of users follows an exponential distribution during training, the generalization performance of both RGNN and 3D-GNN improves, but there still remains a significant gap when compared to the proposed GNN.

\begin{figure}
	\centering
	\vspace{-3mm}
	\includegraphics[width=0.45\textwidth]{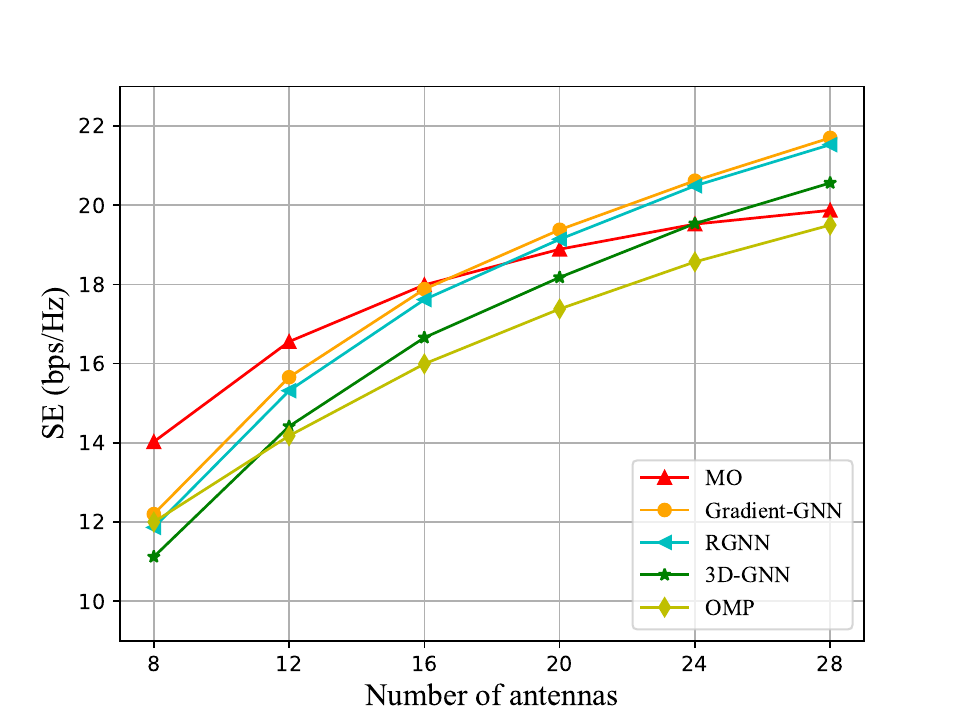}
	\vspace{-2mm}
	\caption{Generalization performance to the number of antennas with $N_s=4$ and $K=4$ (trained with $N=16$).}
	\label{fig:antG-16-4-4}
	\vspace{-3mm}
\end{figure}

In Fig. \ref{fig:antG-16-4-4}, we evaluate the generalization performance with respect to the number of antennas, where the number of antennas $K$ is 4 and the number of RF chains $N_s$ is 4. During the training phase, the number of antennas is set to $N=16$, while during the inference phase, the number of antennas ranges from $8$ to $28$. The results show that all methods exhibit good generalizability to the number of antennas, but the proposed GNN achieves the highest SE compared to other learning methods.

\subsubsection{\underline{Complexity Comparison}}
In this subsection, we compare the training complexity of the GNNs, where \(K=4\), \(N_s=4\) and \(N=16\). The target performance level is set to 90\% of the performance of MO. Inference time is measured as the average runtime over 1000 test samples during the inference stage. The results are summarized in Table~\ref{table: cplxty3} and Table~\ref{table: cplxty5}.

\begin{table}[htb!]
	\centering
	\vspace{-3mm}
	\caption{Training complexity.} \label{table: cplxty3}
	\vspace{-3mm}
	\footnotesize
	\renewcommand\arraystretch{1.1}
	\begin{tabular}{c|c|c|c|c}
		\hline
		Complexity & Grad-GNN & RGNN & 3D-GNN & CHP-ULGAT  \\
		\hline
		Space & 375,556 & 625,032 & 1,319,936 & $>$3,163,296  \\
		\hline
		Sample & 3,000 & 5,000 & 20,000 & $>$100,000 \\
		\hline
            Time (GPU) & 3,016 s & 4,560 s & 3,060 s & $>$6,200 s \\
		\hline
	\end{tabular}
	\vspace{-3mm}
\end{table}

\begin{table}[htb!]
	\centering
	\vspace{-3mm}
	\caption{Inference time.} \label{table: cplxty5}
	\vspace{-3mm}
	\footnotesize
	\renewcommand\arraystretch{1.1}
        \setlength\tabcolsep{2pt}
	\begin{tabular}{c|c|c|c|c|c|c}
		\hline
            
		~& Grad-GNN & RGNN & 3D-GNN & CHP-ULGAT & MO & OMP \\
            \hline
		GPU & 5.92 ms & 6.98 ms & 4.96 ms & $>$2.99 ms & - & - \\
		\hline
		CPU & 24.93 ms & 38.89 ms & 9.97 ms & $>$7.94 ms & 1581 ms & 29.06 ms \\
		\hline
	\end{tabular}
	\vspace{-3mm}
\end{table}


It can be found that the proposed GNN exhibits lower space complexity, sample complexity, and time complexity compared to RGNN and 3D-GNN. The proposed GNN also has lower inference time than RGNN, OMP and MO, although it is slightly higher than that of 3D-GNN. The CHP-ULGAT has very high complexity, achieving only 64.8\% of the performance of MO, even when trained with 100,000 samples.

\section{Conclusions} \label{conclusion}
In this paper, we proposed a gradient-driven GNN design method for precoder learning. By exploring the connection between the gradient descent iteration equation and the GNN update equation, we designed a novel GNN update equation and designed an attention mechanism to learn the aggregation weights.
We demonstrated the versatility of the proposed method by applying it to learn the digital precoders for SE and log-SE maximization, respectively, and to learn the hybrid precoder for SE maximization.
Simulation results showed that the proposed GNN outperforms baseline GNNs in terms of both learning and generalization performance.


\bibliography{IEEEabrv,ZL}

\begin{thebibliography}{10}
\providecommand{\url}[1]{#1}
\csname url@samestyle\endcsname
\providecommand{\newblock}{\relax}
\providecommand{\bibinfo}[2]{#2}
\providecommand{\BIBentrySTDinterwordspacing}{\spaceskip=0pt\relax}
\providecommand{\BIBentryALTinterwordstretchfactor}{4}
\providecommand{\BIBentryALTinterwordspacing}{\spaceskip=\fontdimen2\font plus
\BIBentryALTinterwordstretchfactor\fontdimen3\font minus \fontdimen4\font\relax}
\providecommand{\BIBforeignlanguage}[2]{{%
\expandafter\ifx\csname l@#1\endcsname\relax
\typeout{** WARNING: IEEEtran.bst: No hyphenation pattern has been}%
\typeout{** loaded for the language `#1'. Using the pattern for}%
\typeout{** the default language instead.}%
\else
\language=\csname l@#1\endcsname
\fi
#2}}
\providecommand{\BIBdecl}{\relax}
\BIBdecl

\bibitem{WMMSE}
Q.~Shi, M.~Razaviyayn, Z.-Q. Luo, and C.~He, ``An iteratively weighted {MMSE} approach to distributed sum-utility maximization for a {MIMO} interfering broadcast channel,'' \emph{IEEE Trans. Signal Process.}, vol.~59, no.~9, pp. 4331--4340, Sept. 2011.

\bibitem{DNN1}
H.~Sun, X.~Chen, Q.~Shi, M.~Hong, X.~Fu, and N.~D. Sidiropoulos, ``Learning to optimize: Training deep neural networks for interference management,'' \emph{IEEE Trans. Signal Process.}, vol.~66, no.~20, pp. 5438--5453, Oct. 2018.

\bibitem{DNN2}
J.~Kim, H.~Lee, S.-E. Hong, and S.-H. Park, ``Deep learning methods for universal {MISO} beamforming,'' \emph{IEEE Wireless Commun. Lett.}, vol.~9, no.~11, pp. 1894--1898, Nov. 2020.

\bibitem{DNN3}
K.~Kong, W.-J. Song, and M.~Min, ``Knowledge distillation-aided end-to-end learning for linear precoding in multiuser {MIMO} downlink systems with finite-rate feedback,'' \emph{IEEE Trans. Veh. Tech.}, vol.~70, no.~10, pp. 11\,095--11\,100, Oct. 2021.

\bibitem{DNN6}
Z.~Liu, Y.~Yang, F.~Gao, T.~Zhou, and H.~Ma, ``Deep unsupervised learning for joint antenna selection and hybrid beamforming,'' \emph{IEEE Trans. Wireless Commun.}, vol.~70, no.~3, pp. 1697--1710, March 2022.

\bibitem{DNN7}
T.~Peken, S.~Adiga, R.~Tandon, and T.~Bose, ``Deep learning for {SVD} and hybrid beamforming,'' \emph{IEEE Trans. Wireless Commun.}, vol.~19, no.~10, pp. 6621--6642, Oct. 2020.

\bibitem{SCJ}
C.~Sun, J.~Wu, and C.~Yang, ``Improving learning efficiency for wireless resource allocation with symmetric prior,'' \emph{IEEE Wireless Commun.}, vol.~29, no.~2, pp. 162--168, April 2022.

\bibitem{ZBC}
B.~Zhao, J.~Guo, and C.~Yang, ``Understanding the performance of learning precoding policies with graph and convolutional neural networks,'' \emph{IEEE Trans. Commun.}, vol.~72, no.~9, pp. 5657--5673, Sept. 2024.

\bibitem{PE}
M.~Eisen and A.~Ribeiro, ``Optimal wireless resource allocation with random edge graph neural networks,'' \emph{IEEE Trans. Signal Process.}, vol.~68, pp. 2977--2991, April 2020.

\bibitem{GJ}
J.~Guo and C.~Yang, ``Learning power allocation for multi-cell-multi-user systems with heterogeneous graph neural networks,'' \emph{IEEE Trans. Wireless Commun.}, vol.~21, no.~2, pp. 884--897, Feb. 2022.

\bibitem{LSJ}
S.~Liu, J.~Guo, and C.~Yang, ``Learning hybrid precoding efficiently for mm{Wave} systems with mathematical properties,'' in \emph{Proc. IEEE GLOBECOM}, 2022.

\bibitem{MDGNN}
------, ``Multidimensional graph neural networks for wireless communications,'' \emph{IEEE Trans. Wireless Commun.}, vol.~23, no.~4, pp. 3057--3073, April 2024.

\bibitem{WMMSE1}
Q.~Hu, Y.~Cai, Q.~Shi, K.~Xu, G.~Yu, and Z.~Ding, ``Iterative algorithm induced deep-unfolding neural networks: Precoding design for multiuser {MIMO} systems,'' \emph{IEEE Trans. Wireless Commun.}, vol.~20, no.~2, pp. 1394--1410, Feb. 2021.

\bibitem{WMMSE2}
A.~Chowdhury, G.~Verma, C.~Rao, A.~Swami, and S.~Segarra, ``Unfolding {WMMSE} using graph neural networks for efficient power allocation,'' \emph{IEEE Trans. Wireless Commun.}, vol.~20, no.~9, pp. 6004--6017, Sept. 2021.

\bibitem{WMMSE3}
A.~Chowdhury, G.~Verma, A.~Swami, and S.~Segarra, ``Deep graph unfolding for beamforming in {MU-MIMO} interference networks,'' \emph{IEEE Trans. Wireless Commun.}, vol.~23, no.~5, pp. 4889--4903, May 2024.

\bibitem{WMMSE4}
L.~Schynol and M.~Pesavento, ``Coordinated sum-rate maximization in multicell {MU-MIMO} with deep unrolling,'' \emph{IEEE J. Sel. Areas Commun.}, vol.~41, no.~4, pp. 1120--1134, April 2023.

\bibitem{WMMSE5}
J.~Zhang, C.~Masouros, and L.~Hanzo, ``Joint precoding and {CSI} dimensionality reduction: An efficient deep unfolding approach,'' \emph{IEEE Trans. Wireless Commun.}, vol.~22, no.~12, pp. 9502--9516, Dec. 2023.

\bibitem{optimal1}
W.~Xia, G.~Zheng, Y.~Zhu, J.~Zhang, J.~Wang, and A.~P. Petropulu, ``A deep learning framework for optimization of {MISO} downlink beamforming,'' \emph{IEEE Trans. Commun.}, vol.~68, no.~3, pp. 1866--1880, March 2020.

\bibitem{optimal2}
H.~Huang, Y.~Peng, J.~Yang, W.~Xia, and G.~Gui, ``Fast beamforming design via deep learning,'' \emph{IEEE Trans. Veh. Tech.}, vol.~69, no.~1, pp. 1065--1069, Jan. 2020.

\bibitem{optimal3}
Y.~Yuan, G.~Zheng, K.-K. Wong, B.~Ottersten, and Z.-Q. Luo, ``Transfer learning and meta learning-based fast downlink beamforming adaptation,'' \emph{IEEE Trans. Wireless Commun.}, vol.~20, no.~3, pp. 1742--1755, March 2021.

\bibitem{optimal4}
J.~Kim, H.~Lee, S.-E. Hong, and S.-H. Park, ``Deep learning methods for universal {MISO} beamforming,'' \emph{IEEE Wireless Commun. Lett.}, vol.~9, no.~11, pp. 1894--1898, Nov. 2020.

\bibitem{optimal5}
------, ``A bipartite graph neural network approach for scalable beamforming optimization,'' \emph{IEEE Trans. Wireless Commun.}, vol.~22, no.~1, pp. 333--347, Jan. 2023.

\bibitem{DR}
J.~Guo and C.~Yang, ``Deep neural networks with data rate model: Learning power allocation efficiently,'' \emph{IEEE Trans. Commun.}, vol.~71, no.~3, pp. 1447--1461, March 2023.

\bibitem{model}
N.~Shlezinger, J.~Whang, Y.~C. Eldar, and A.~G. Dimakis, ``Model-based deep learning,'' \emph{Proceedings of the IEEE}, vol. 111, no.~5, pp. 465--499, May 2023.

\bibitem{DNN5}
F.~Sohrabi, K.~M. Attiah, and W.~Yu, ``Deep learning for distributed channel feedback and multiuser precoding in {FDD} massive {MIMO},'' \emph{IEEE Trans. Wireless Commun.}, vol.~20, no.~7, pp. 4044--4057, July 2021.

\bibitem{DNN4}
T.~Jiang, H.~V. Cheng, and W.~Yu, ``Learning to reflect and to beamform for intelligent reflecting surface with implicit channel estimation,'' \emph{IEEE J. Sel. Areas Commun.}, vol.~39, no.~7, pp. 1931--1945, July 2021.

\bibitem{DetNet}
N.~Samuel, T.~Diskin, and A.~Wiesel, ``Learning to detect,'' \emph{IEEE Trans. Signal Process.}, vol.~67, no.~10, pp. 2554--2564, May 2019.

\bibitem{OAMPNet}
H.~He, C.-K. Wen, S.~Jin, and G.~Y. Li, ``Model-driven deep learning for {MIMO} detection,'' \emph{IEEE Trans. Signal Process.}, vol.~68, pp. 1702--1715, Feb. 2020.

\bibitem{EPGNN}
A.~Kosasih, V.~Onasis, V.~Miloslavskaya, W.~Hardjawana, V.~Andrean, and B.~Vucetic, ``Graph neural network aided {MU-MIMO} detectors,'' \emph{IEEE J. Sel. Areas Commun.}, vol.~40, no.~9, pp. 2540--2555, Sept. 2022.

\bibitem{AMPGNN}
H.~He, X.~Yu, J.~Zhang, S.~Song, and K.~B. Letaief, ``Message passing meets graph neural networks: A new paradigm for massive {MIMO} systems,'' \emph{IEEE Trans. Wireless Commun.}, vol.~23, no.~5, pp. 4709--4723, May 2024.

\bibitem{catDHD}
J.~Guo and C.~Yang, ``How to improve learning efficiency of {GNN} for precoding?'' in \emph{Proc. IEEE VTC Spring}, 2023.

\bibitem{ZL-GCW23}
L.~Zhang, S.~Han, C.~Yang, and Y.~Li, ``A gradient driven graph neural network for optimizing precoding,'' in \emph{Proc. IEEE GLOBECOM Workshops}, 2023.

\bibitem{GAT}
P.~Veli{\v{c}}kovi{\'c}, G.~Cucurull, A.~Casanova, A.~Romero, P.~Li{\`o}, and Y.~Bengio, ``Graph attention networks,'' in \emph{Proc. ICLR}, 2018.

\bibitem{WMMSE-PF}
M.~Zaher, {\"O}.~T. Demir, E.~Bj{\"o}rnson, and M.~Petrova, ``Learning-based downlink power allocation in cell-free massive {MIMO} systems,'' \emph{IEEE Trans. Wireless Commun.}, vol.~22, no.~1, pp. 174--188, Jan. 2023.

\bibitem{MO}
X.~Yu, J.-C. Shen, J.~Zhang, and K.~B. Letaief, ``Alternating minimization algorithms for hybrid precoding in millimeter wave {MIMO} systems,'' \emph{IEEE J. Sel. Top. Signal Process}, vol.~10, no.~3, pp. 485--500, April 2016.

\bibitem{OMP}
O.~E. Ayach, S.~Rajagopal, S.~Abu-Surra, Z.~Pi, and R.~W. Heath, ``Spatially sparse precoding in millimeter wave {MIMO} systems,'' \emph{IEEE Trans. Wireless Commun.}, vol.~13, no.~3, pp. 1499--1513, March 2014.

\bibitem{RGNN}
J.~Guo and C.~Yang, ``A size-generalizable {GNN} for learning precoding,'' in \emph{Proc. IEEE VTC Fall}, 2023.

\bibitem{GAT+FCNN}
Y.~Zhang, J.~Yang, Q.~Liu, Y.~Liu, and T.~Zhang, ``Unsupervised learning-based coordinated hybrid precoding for {mmWave} massive {MIMO}-enabled {HetNets},'' \emph{IEEE Trans. Wireless Commun.}, vol.~23, no.~7, pp. 7200--7213, July 2024.

\end{thebibliography}








\end{document}